\documentclass{svmult}

\usepackage{makeidx}
\usepackage{graphicx}
\usepackage{multicol}

\def\beq{\begin{equation}}
\def\eeq{\end{equation}}
\def\bea{\begin{eqnarray}}
\def\eea{\end{eqnarray}}

\begin{document}

\title*{Hawking radiation from higher-dimensional black holes}

\titlerunning{Hawking radiation from higher-dimensional black holes}

\author{Panagiota Kanti\inst{1}\and
Elizabeth Winstanley\inst{2}}

\institute{
Division of Theoretical Physics,
Department of Physics,
University of Ioannina,\\
Ioannina. GR-451 10
Greece\\
\texttt{pkanti@cc.uoi.gr}
\and
Consortium for Fundamental Physics,
School of Mathematics and Statistics,\\
The University of Sheffield,
Hicks Building, Hounsfield Road,
Sheffield. S3 7RH
United Kingdom\\
\texttt{e.winstanley@sheffield.ac.uk}}

\maketitle

\abstract{
We review the quantum field theory description of Hawking
radiation from evaporating black holes and summarize what
is known about Hawking radiation from black holes in more than
four space-time dimensions. In the context of the Large Extra
Dimensions scenario, we present the theoretical formalism for
all types of emitted fields and a selection of results
on the radiation spectra. A detailed analysis of the Hawking
fluxes in this case is essential for modelling the evaporation of
higher-dimensional black holes at the LHC, whose creation is
predicted by low-energy models of quantum gravity. We discuss the
status of the quest for black-hole solutions in the context of the
Randall-Sundrum brane-world model and, in the absence of an exact
metric, we review what is known about Hawking radiation from
such black holes.
}

\section{Introduction}
\label{sec:intro}

Hawking radiation \cite{Hawking:1975sw} is one of the most important
effects arising from quantum field theory in curved space, a semi-classical
approach to quantum gravity.
In this framework space-time is described by a classical geometry, governed by
the Einstein equations (or an alternative classical theory of gravity).
The behaviour and propagation of quantum fields on a fixed (but not necessarily
stationary) space-time is then studied.
Hawking radiation is thermal in nature, giving non-extremal black holes an intrinsic temperature
proportional to the surface gravity of the event horizon.
For a Schwarzschild black hole in asymptotically flat space, the specific heat is negative, so that the temperature
increases as the black hole evaporates, leading to a black hole explosion.
The ultimate fate of the black hole depends on unknown details of quantum gravity, but black hole evaporation
raises many deep questions about the nature of quantum gravity and the fundamental laws of physics
(such as the information loss paradox, see, for example, \cite{Hossenfelder:2009xq,Mathur:2009hf}).
Here we will not address these important issues, but instead focus on the detailed properties of the Hawking radiation itself.

Hawking radiation from four-dimensional black holes in asymptotically flat space was studied in detail by Page
\cite{Page:1976df,Page:1976ki,Page:1977um}.
Over the past fifteen or so years,  there has been great interest in higher-dimensional black holes.
Within the context of classical general relativity, a menagerie of black-hole-like solutions of the Einstein equations has been
discovered (see for example \cite{Emparan:2008eg,Tomizawa:2011mc} for reviews).
It is then natural to study the properties of Hawking radiation from higher-dimensional black holes.

This avenue of research gained much impetus from the exciting possibility of producing microscopic higher-dimensional black holes
in high-energy collisions either at the LHC or in cosmic rays
\cite{Banks:1999gd,Dimopoulos:2001hw,Giddings:2001bu, Kanti:2004nr,Landsberg:2003br,Majumdar:2005ba,Park:2012fe,Webber:2005qa}.
This is a prediction of higher-dimensional brane-world models \cite{ArkaniHamed:1998rs,Antoniadis:1998ig,ArkaniHamed:1998nn,Randall:1999ee,Randall:1999vf}
in which the energy scale of quantum gravity is much lower than the traditional value of $10^{19}$ GeV, and may be as low as the TeV-scale.
If such a microscopic black hole is produced, it will initially be rapidly rotating and rather asymmetric.
Its subsequent evolution can be modelled as four stages \cite{Giddings:2001bu}:
\begin{itemize}
\item
During the {\em {balding phase}} the black hole sheds its asymmetries through the emission of gravitational radiation and also loses any
gauge field charges arising from the particles which formed it. At the end of this stage the black hole is axisymmetric and still
rapidly rotating.
\item
The black hole then emits Hawking radiation, and loses both mass and angular momentum.  At the end of this {\em {spin-down phase}}
the black hole is no longer rotating.
\item
Now with zero angular momentum, the black hole continues to radiate during the {\em {Schwarzschild phase}}, shrinking as it loses mass.
\item
During the final {\em {Planck phase}} the semi-classical approximation for the Hawking radiation is no longer valid and the black hole emission depends on the details of quantum gravity.
\end{itemize}
It is expected that the spin-down and Schwarzschild phases will dominate the life-time of the black hole.
A detailed understanding of the Hawking radiation from higher-dimensional black holes in brane-world models is therefore necessary
for simulating microscopic black hole events \cite{Dai:2007ki,Frost:2009cf} and experimental searches, as well as being of intrinsic
theoretical interest.

In this chapter we focus on the theoretical modelling of Hawking radiation from higher-dimensional black holes.
We begin with a discussion of the quantum-field-theoretic derivation of Hawking radiation and its description using the Unruh vacuum state
\cite{Unruh:1976db}.
We then briefly review some key features of black holes in brane world models.
We describe the formalism for studying quantum fields on higher-dimensional Myers-Perry black holes \cite{Myers:1986un}, which
model black holes in an ADD brane-world \cite{ArkaniHamed:1998rs,Antoniadis:1998ig,ArkaniHamed:1998nn}.
We also present a selection of results on the properties of the Hawking radiation from these black holes.
The literature on this subject is vast and so we cannot claim to do all aspects justice.
The reviews \cite{Kanti:2004nr,Casanova:2005id,Kanti:2008eq,Kanti:2012jh,Winstanley:2007hj} contain further discussion of results which space does not
permit us to include.
In the RS brane-world \cite{Randall:1999ee,Randall:1999vf}, analysis of the Hawking radiation is more challenging because no exact metric describing a five-dimensional black hole localised on the brane
is known in general - for a more detailed discussion of this topic, see the reviews
\cite{Gregory_Aegean, Kanti:2009sz, Tanahashi}.
We close the chapter with a discussion of what is known about the Hawking radiation in this case.

\section{Hawking radiation}
\label{sec:Hawkrad}

\subsection{Hawking radiation from a black hole formed by gravitational collapse}
\label{sec:gravcoll}

\begin{figure}
\begin{center}
\includegraphics[width=3cm]{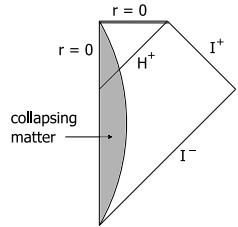}
\end{center}
\caption{Penrose diagram for a Schwarzschild black hole formed by gravitational collapse.}
\label{fig:collapse}
\end{figure}

\begin{figure}
\begin{center}
\includegraphics[width=6cm]{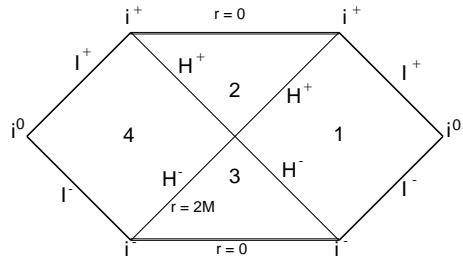}
\end{center}
\caption{Penrose diagram for an eternal Schwarzschild black hole}
\label{fig:schwarzschild}
\end{figure}

Hawking's original derivation \cite{Hawking:1975sw} considered a quantum scalar field
propagating on a fixed, but dynamic, background space-time corresponding to the
formation of a four-dimensional Schwarzschild black hole by the
gravitational collapse of matter in asymptotically flat space.
The Penrose diagram for this process is shown in Figure~\ref{fig:collapse}
(cf. the Penrose diagram for an eternal Schwarzschild black hole in Figure~\ref{fig:schwarzschild}).


For the moment consider a massless scalar field in a two-dimensional version of the space-time
shown in Figure~\ref{fig:collapse}, with space-time co-ordinates $(t,r)$.
In this case, because two-dimensional space-times are locally conformally flat,
a basis of field modes is formed simply from plane waves.
These plane waves are not normalizable, but normalizable wave-packets can be constructed from appropriate linear combinations of the plane wave modes.
At very early times, long before the collapse starts, a suitable basis of field modes is:
\begin{equation}
\phi _{\omega } \propto e^{-i\omega t} e^{\pm i\omega r},
\label{eq:2dmodes}
\end{equation}
where $\omega >0$ corresponds to positive frequency.
The quantum scalar field ${\hat {\Phi }}$ is written in terms of these basis modes:
\begin{equation}
{\hat {\Phi }} = \int _{\omega =0}^{\infty } d\omega \, \left[
 {\hat {a}}_{\omega } \phi _{\omega }
 + {\hat {a}}_{\omega}^{\dagger } \phi ^{*}_{\omega }
\right] ,
\label{eq:2dphi}
\end{equation}
where the expansion coefficients $a_{\omega }$ have been promoted to operators in the canonical quantization of the scalar field.
Working in the Heisenberg picture, the quantum state is defined to be the ``in'' vacuum at early times near $I^{-}$, namely the state
$\left| 0 \right\rangle _{{\mathrm {in}}}$ which is annihilated by the
${\hat {a}}_{\omega }$ operators:
\begin{equation}
{\hat {a}}_{\omega } \left| 0 \right\rangle _{{\mathrm {in}}} =0 \qquad \forall
\omega >0.
\end{equation}
At late times, long after the black hole has formed, we can form a basis of field modes similar to (\ref{eq:2dmodes}), write the quantum scalar field in terms of these modes along the lines of (\ref{eq:2dphi}) but with the expansion coefficients
${\hat {a}}_{\omega }$ now replaced by operators ${\hat {b}}_{\omega }$, and define an ``out'' vacuum state $\left| 0 \right\rangle _{{\mathrm {out}}}$ which is annihilated
by the ${\hat {b}}_{\omega }$ operators.

The crux of Hawking's derivation \cite{Hawking:1975sw} is that the ``in'' and ``out''
vacuum states are not the same: the ``in'' vacuum
$\left| 0 \right\rangle _{{\mathrm {in}}}$ contains a thermal flux of outgoing
particles at late times near $I^{+}$ (we use units in which $c=\hbar = G= k_{B}=1$):
\begin{equation}
{}_{{\mathrm {in}}}\left\langle 0 \left|
{\hat {b}}_{\omega }^{\dagger } {\hat {b}}_{\omega }
\right| 0 \right\rangle _{{\mathrm {in}}} = \frac {1}{e^{\omega /T_{H}}-1}
\label{eq:Hflux1}
\end{equation}
where
\begin{equation}
T_{H}= \frac {\kappa }{2\pi }
\label{eq:temperature}
\end{equation}
is the Hawking temperature and $\kappa $ is the surface gravity of the black hole.
There are many different derivations of this effect (see, for example,
\cite{Hartle:1976tp,Damour:1976jd,Gerlach:1976ji,Parentani:1992me,Brout:1995rd,Hu:1996br,Massar:1999wg,Parikh:1999mf,Schutzhold:2000ju,Shankaranarayanan:2000qv,Visser:2001kq,Peltola:2008jx}). Hawking's result is very robust, and essentially kinematic: it is independent of the Einstein equations or the theory of gravity under consideration.
There are a number of different pictures to understand the origin of the thermal flux, such as quantum tunnelling of classically forbidden trajectories from behind the event horizon \cite{Parikh:1999mf}, or the pair creation of quantum particles close to the event horizon, one of which carries negative energy down the event horizon and the
other of which escapes to infinity.
In terms of modes, the thermal factor arises because outgoing modes at $I^{+}$ can be traced back to ingoing modes which enter the collapsing body just before the last ingoing mode (which then forms the event horizon of the black hole after reflection from the co-ordinate centre $r=0$), resulting in a ``pile-up'' of highly blue-shifted modes near this last ingoing mode.

Using a geometric optics argument, Hawking's result can be extended to black holes in four or more space-time dimensions \cite{Hawking:1975sw}.
The complication is that, even for a massless scalar field, in more than two space-time dimensions the quantum field interacts with a gravitational potential
which surrounds the black hole.
As a result, a wave which is outgoing near the event horizon of the black hole will
partly escape to infinity and partly be reflected back down the event horizon.
The part which escapes to $I^{+}$ will contribute to the Hawking flux.
In four or more space-time dimensions, each mode of a quantum field of spin $s$
will be characterized
by its frequency $\omega $, a total angular momentum quantum number $\ell $,
an azimuthal quantum number $m$ (indexing the angular momentum about the $z$-axis) and, in more than four space-time dimensions, further angular quantum numbers $j$.
To describe this scattering effect we introduce the grey-body factor
$\Gamma _{s\omega \ell m j}$ which is given by the outgoing flux near $I^{+}$ for each mode divided by the outgoing flux near the horizon in that mode (that is, the fraction of each outgoing mode near the horizon which is transmitted to $I^{+}$).
The Hawking flux (\ref{eq:Hflux1}) is then modified to be, for each quantum field mode:
\begin{equation}
{}_{{\mathrm {in}}}\left\langle 0 \left|
{\hat {b}}_{\omega }^{\dagger } {\hat {b}}_{\omega }
\right| 0 \right\rangle _{{\mathrm {in}}} =
\frac {\Gamma _{s\omega \ell m j}}{e^{\omega /T_{H}}\pm 1} ,
\label{eq:Hflux2}
\end{equation}
where the $+$ sign in the denominator is for fermionic fields and the $-$ sign
for bosonic fields.
While Hawking's original derivation \cite{Hawking:1975sw} was for a quantum scalar field, we emphasize that the result carries over to quantum fields of all spins.
Furthermore, although in the above we have considered a Schwarzschild black hole,
any black hole with a non-extremal event horizon will emit Hawking radiation, including rotating black holes.
In this article we will consider only rotating black holes with a single axis of rotation (which we take to be the $z$-axis).
In this case the denominator of the Hawking flux (\ref{eq:Hflux2}) is modified by the rotation of the black hole to be
\begin{equation}
e^{{\tilde {\omega }}/T}\pm 1
\end{equation}
where
\begin{equation}
{\tilde {\omega }}= \omega - m \Omega _{H},
\label{eq:tildeomega}
\end{equation}
with $m$ the azimuthal quantum number and $\Omega _{H}$ is the angular velocity of the event horizon.

\subsection{The Unruh state}
\label{sec:Unruh}

In practice, dealing with quantum fields on the dynamical space-time shown in
Figure~\ref{fig:collapse} is technically difficult.
In computing Hawking radiation, a different approach is usually employed.
Instead of considering the collapse geometry of Figure~\ref{fig:collapse},
the eternal black hole space-time (such as that for the Schwarzschild black hole
shown in Figure~\ref{fig:schwarzschild}) is considered.
We restrict our attention to the right-hand-diamond of Figure~\ref{fig:schwarzschild}, representing the region exterior to the black hole event horizon.
Charged and/or rotating black holes have more complex Penrose diagrams than
Figure~\ref{fig:schwarzschild}, but the diamond-shaped region exterior to the event horizon is the same for all asymptotically-flat black holes.
For black holes in de Sitter space, the relevant region is that between the
event and cosmological horizons, which has the same diamond shape.

\begin{figure}
\begin{center}
\includegraphics[width=3cm]{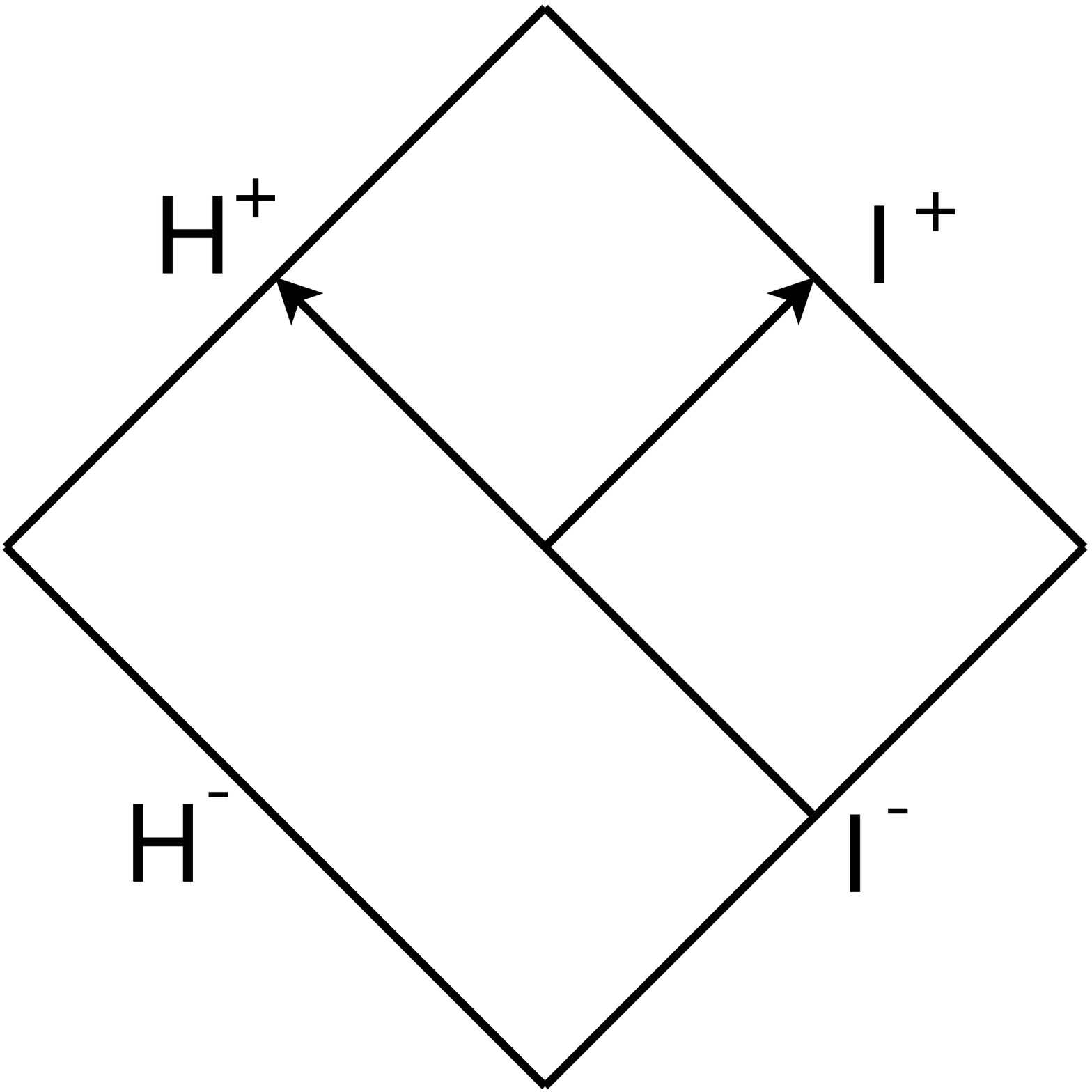}
\, \, \, \,
\includegraphics[width=3cm]{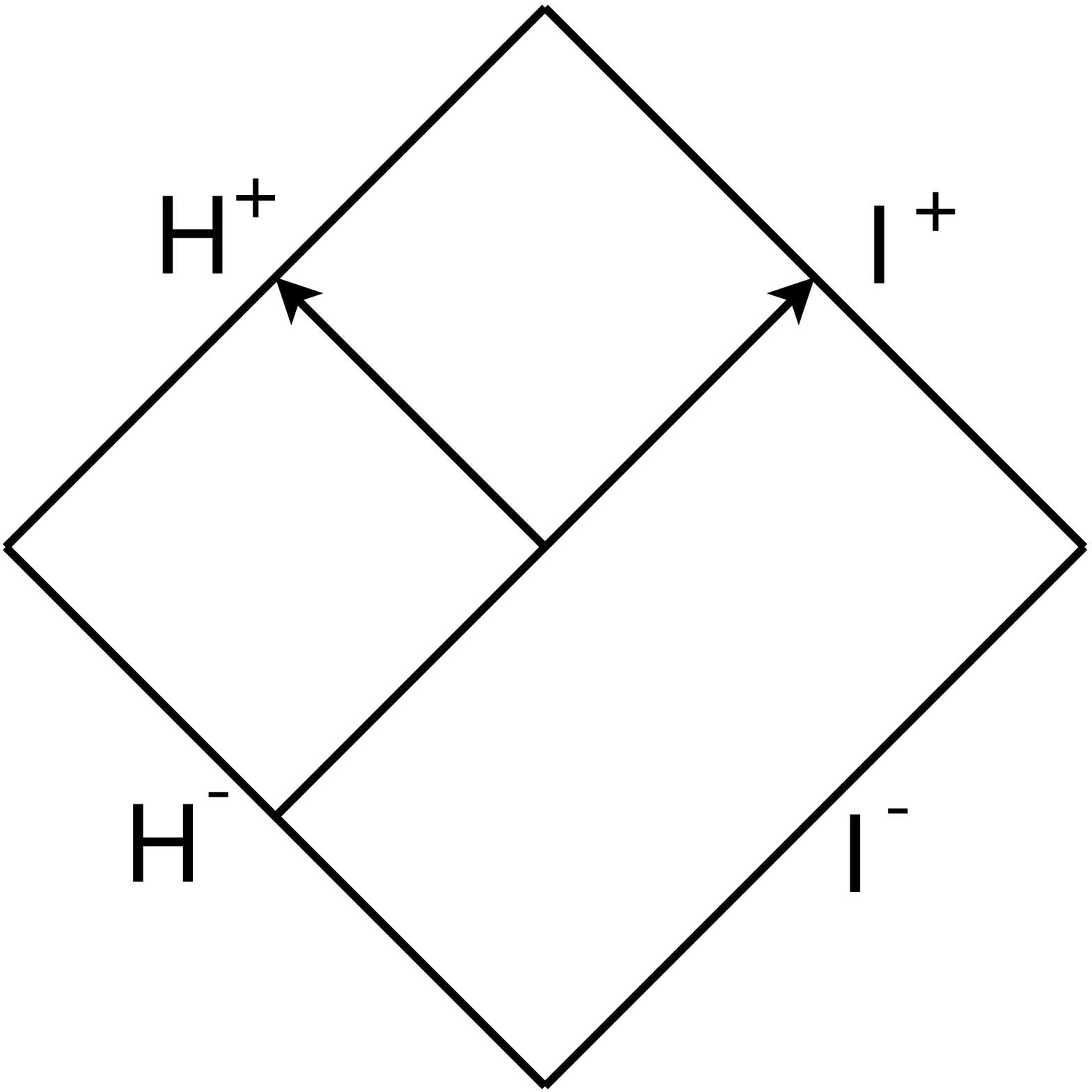}
\end{center}
\caption{``In'' (left) and ``up'' (right) modes}
\label{fig:inup}
\end{figure}

The Unruh state \cite{Unruh:1976db} is the quantum state on the eternal black hole space-time which models the Hawking radiation.
The construction of this state proceeds as follows, where for simplicity we consider
a free massless scalar field $\Phi $ and a four-dimensional Schwarzschild black hole
with co-ordinates $(t, r, \theta , \varphi )$.
In this case the scalar field modes $\phi _{\omega \ell m}$ are separable:
\begin{equation}
\phi _{\omega \ell m}(t,r, \theta ,\varphi ) = e^{-i\omega t}e^{im\varphi } R_{\omega \ell m}(r)
Y_{\ell m}(\theta ),
\label{eq:simplemodes}
\end{equation}
where $\omega $ is the frequency of the mode, $m$ is the azimuthal quantum number,
$\ell $ is the total-angular-momentum quantum number, $R_{\omega \ell m}(r)$ is the radial function and $Y_{\ell m}(\theta )$ is a spherical harmonic.
First a basis of quantum field modes is required.
We take as a basis the ``in'' and ``up'' modes depicted in Figure~\ref{fig:inup}.
The ``in'' modes $\phi _{\omega \ell m}^{{\mathrm {in}}}$ are incoming from past null infinity $I^{-}$. Part of each wave
is reflected by the gravitational potential and scatters back to infinity, and part
goes down the future event horizon $H^{+}$.
The ``up'' modes $\phi _{\omega \ell m}^{{\mathrm {up}}}$ are outgoing from close to the past event horizon $H^{-}$.
In this case part of each wave is reflected back down the future event horizon
$H^{+}$, while part is transmitted to future null infinity $I^{+}$.

Having chosen a suitable basis of field modes, we now need to split these modes into
positive and negative frequency.
For the ``in'' modes, we choose `positive frequency' to mean `positive frequency as seen by a static observer far from the black hole', so that, for a field mode
$\phi _{\omega \ell m}$ we have:
\begin{equation}
\frac {\partial }{\partial t} \phi _{\omega \ell m}= -i\omega \phi _{\omega \ell m},
\end{equation}
where $\omega >0$.
Since the ``in'' modes originate from near $I^{-}$, this is the most natural choice
of positive frequency for these modes.
For the ``up'' modes, we choose `positive frequency' to mean `positive frequency with respect to Kruskal time near the event horizon'.
Since the ``up'' modes originate from near $H^{-}$, this choice of positive frequency, with respect to co-ordinates which are regular across the event horizon, is also very natural.
We decompose our quantum field ${\hat {\Phi }}$ in terms of this basis of positive frequency modes:
\begin{equation}
{\hat {\Phi }}= \sum _{{\mathrm {modes}}}
\left[
{\hat {a}}_{\omega \ell m}^{{\mathrm {in}}} \phi _{\omega \ell m}^{{\mathrm {in}}}
+
{\hat {a}}_{\omega \ell m}^{{\mathrm {in}}\dagger }
\phi _{\omega \ell m}^{{\mathrm {in}} *}
\right]
+
\sum _{{\mathrm {modes}}}
\left[
{\hat {a}}_{\omega \ell m}^{{\mathrm {up}}} \phi _{\omega \ell m}^{{\mathrm {up}}}
+
{\hat {a}}_{\omega \ell m}^{{\mathrm {up}}\dagger } \phi _{\omega \ell m}^{{\mathrm {up}}*}
\right]
\end{equation}
where in each case the sum is taken over the positive frequency modes and we have promoted the expansion coefficients for the classical scalar field to operators satisfying the usual commutation relations:
\begin{equation}
\left[ {\hat {a}}_{\omega \ell m}^{{\mathrm {in/up}}},
{\hat {a}}_{\omega '\ell 'm'}^{{\mathrm {in/up}}\dagger } \right]
= \delta (\omega - \omega ') \delta _{\ell ,\ell '} \delta _{m,m'},
\quad
\left[ {\hat {a}}_{\omega \ell m}^{{\mathrm {in/up}}},
{\hat {a}}_{\omega '\ell 'm'}^{{\mathrm {in/up}}} \right]
= 0 =
\left[ {\hat {a}}_{\omega \ell m}^{{\mathrm {in/up}}\dagger },
{\hat {a}}_{\omega '\ell 'm'}^{{\mathrm {in/up}}\dagger } \right] .
\end{equation}
The Unruh state $\left| U\right\rangle $ \cite{Unruh:1976db} is then defined as
that state which is annihilated by the operators
${\hat {a}}_{\omega \ell m}^{{\mathrm {in/up}}}$:
\begin{equation}
{\hat {a}}_{\omega \ell m}^{{\mathrm {in/up}}}\left| U \right\rangle =0.
\end{equation}
This state has no particles in the ``in'' modes near $I^{-}$ as was the case for the
``in'' vacuum $\left| 0 \right\rangle _{{\mathrm {in}}}$ describing the state for a
black hole formed by gravitational collapse.
However, due to the choice of positive frequency for the ``up'' modes, these modes
are thermally populated with temperature $T_{H}$ (\ref{eq:temperature}).
Therefore, near future null infinity $I^{+}$, a static observer will see an outgoing flux of particles in the ``up'' modes, which is precisely the Hawking radiation.
Furthermore, the flux in each mode will be given by (\ref{eq:Hflux2}), with the
grey-body factor $\Gamma _{\omega \ell m}$ (with spin $s=0$ and no index $j$ because we are in four space-time dimensions) representing the
proportion of each ``up'' mode which escapes to $I^{+}$.

Here we have discussed the construction of the Unruh state for the particular example of a massless scalar field on a four-dimensional Schwarzschild space-time.
The extension of this construction to higher-dimensional, spherically-symmetric space-times is straightforward.
For higher-spin fields, the field modes are more complicated than (\ref{eq:simplemodes}) but the construction works in a similar way
(bearing in mind that the Hawking flux (\ref{eq:Hflux2}) has a $+$
sign in the denominator for fermion fields and a $-$ sign for bosonic fields).
The Unruh state can also be constructed for rotating black holes.
In this case the frequency $\omega $ in the thermal factor in (\ref{eq:Hflux2})
becomes ${\tilde {\omega }}$ (\ref{eq:tildeomega}).
This shift arises because ${\tilde {\omega }}$ rather than $\omega $ is the natural
frequency of field modes near the horizon of a rotating black hole, from where the
Hawking radiation emanates.
For details of this construction on four-dimensional Kerr black holes, see
\cite{Ottewill:2000qh} (scalars), \cite{Casals:2012es} (fermions) and
\cite{Casals:2005kr} (electromagnetism).

In order to compute the fluxes of particles, energy and angular momentum emitted in Hawking radiation we calculate expectation values of the appropriate operators in the Unruh state on an eternal black hole space-time.
The flux of particles per unit time is given by summing the individual mode flux (\ref{eq:Hflux2})
over all field modes:
\begin{equation}
\frac {dN}{dt} = \sum _{{\mathrm {modes}}}
\frac {\Gamma _{s\omega \ell m j}}{e^{{\tilde {\omega }}/T_{H}}\pm 1} ,
\label{eq:simpleparticleflux}
\end{equation}
where the exact form of the sum over the modes will be made precise in section~\ref{sec:MPrad}, and ${\tilde {\omega }}=\omega $ if the black hole is non-rotating.
The fluxes of energy and (for a rotating black hole) angular momentum per unit time
are given by expectation values of the quantum stress-energy tensor for the particular quantum field under consideration:
\begin{equation}
\frac {dE}{dt} = \int r^{2}\, d\Omega \left\langle U \left| {\hat {T}}^{r}_{t}
\right| U \right\rangle ,
\qquad
\frac {dJ}{dt} = \int r^{2}\, d\Omega  \left\langle U \left| {\hat {T}}^{r\varphi }
\right| U \right\rangle ,
\label{eq:simplefluxes}
\end{equation}
where the integral is taken over the sphere at infinity and we have given the expressions for four-dimensional black holes.

Expectation values of the stress-energy tensor typically require renormalization, due to their involving products of two operators at the same space-time point.
One method of renormalization is covariant geodesic point separation, in which the two operators whose products are taken are evaluated at different space-time points $x$
and $x'$, yielding a finite bitensor stress-energy tensor, whose expectation value is written $T_{\mu \nu }(x,x')$.
This expectation value is renormalized by the subtraction of purely geometric, state
independent renormalization terms
$T^{{\mathrm {div}}}_{\mu \nu }(x,x')$ (see for example
\cite{Christensen:1976vb,Christensen:1978yd} for expressions for these geometric terms for fields of spin $0$, $1/2$ and $1$ in four dimensions, and
\cite{Decanini:2005eg} for a scalar field in higher-dimensional space-time).
The points $x$ and $x'$ are then brought together and a finite renormalized expectation value for the stress-energy tensor is yielded.

Fortunately, for the black holes in which we are interested, it can be shown that the
two stress-energy tensor components in (\ref{eq:simplefluxes}) do not require renormalization.
For a quantum scalar field on a four-dimensional Kerr black hole, this was shown by Frolov and Thorne \cite{Frolov:1989jh}.
Their argument involved two key properties: (i) the symmetry of the underlying
space-time under the reflection $\left( t, \varphi \right)  \rightarrow
\left( -t, -\varphi \right) $ (simultaneous reversal of time and the azimuthal angle);
and (ii) each of the geometric subtraction terms involves an even number of covariant derivatives $\sigma ^{\mu }$ of the biscalar of geodetic interval \cite{J.L.Synge:1960zz} when an average has been taken over a separation in the
$\sigma ^{\mu }$ and $-\sigma ^{\mu }$ directions.
Choosing radial point-splitting, these two properties ensure that the geometric
subtraction terms $T_{tr}^{{\mathrm {div}}}(x,x')$ and
$T_{r\varphi }^{{\mathrm {div}}}(x,x')$ both vanish.
Since the two properties above are shared by the simply-rotating black holes which
we shall consider in section~\ref{sec:MPrad}, the above argument can be readily extended to show that $T_{tr}^{{\mathrm {div}}}(x,x')$ and
$T_{r\varphi }^{{\mathrm {div}}}(x,x')$ both vanish for these black holes as well
\cite{Casals:2008pq}.

The exact form of the stress-energy tensor components in (\ref{eq:simplefluxes})
depends on the spin of the quantum field under consideration
(see, for example, \cite{Kanti:2004nr,Casals:2008pq,Duffy:2005ns,Casals:2006xp,Casals:2005sa}
for details).
However, the resulting fluxes of energy and angular momentum
have the following simple forms:
\begin{equation}
\frac {dE}{dt} = \sum _{{\mathrm {modes}}}
\frac {\omega \Gamma _{s\omega \ell m j}}{e^{{\tilde {\omega }}/T_{H}}\pm 1},
\qquad
\frac {dJ}{dt} = \sum _{{\mathrm {modes}}}
\frac {m \Gamma _{s\omega \ell m j}}{e^{{\tilde {\omega }}/T_{H}}\pm 1},
\label{eq:simplefluxes1}
\end{equation}
where $m$ is the azimuthal quantum number.
More precise details of the mode sums can be found in section~\ref{sec:greybody}.

\section{Brane world black holes}
\label{sec:BBH}

\subsection{Black holes in ADD brane-worlds}
\label{sec:ADDbh}

In the ADD brane-world scenario \cite{ArkaniHamed:1998rs,Antoniadis:1998ig,
ArkaniHamed:1998nn} space-time has $d=4+n$ dimensions.
Our universe is a four-dimensional brane in this higher-dimensional bulk space-time.
The $n$ extra dimensions are flat and compactified (the radius of compactification is
typically large compared with the Planck length but sufficiently
small to agree with searches for deviations from Newton's Law of Gravitation).
To avoid contradictions with precision particle-physics experiments, the forces and
particles of the Standard Model are constrained to live on the brane;
only gravitational degrees of freedom (gravitons and possibly scalars) can propagate
in the bulk.

We model black holes in the ADD scenario in a very simple way, assuming that the brane
is tensionless and infinitely thin. We also assume that the black-hole horizon is much
smaller than the compactification radius of the extra dimensions. Effectively we are
considering black holes in an asymptotically flat, $(4+n)$-dimensional, space-time.
Furthermore, we are particularly interested in microscopic black holes formed by the collision of particles on
the brane (such a collision will not necessarily be head-on). In this case, by
conservation of angular momentum, the resulting black hole will have a single axis
of rotation, which will also lie in the brane.

Rotating black-hole solutions of the vacuum Einstein equations in $(4+n)$-dimensional
space-time are described by the Myers-Perry metric \cite{Myers:1986un}.
Unlike the situation in four space-time dimensions, the Myers-Perry metric is not
unique if $n>0$ \cite{Emparan:2008eg,Tomizawa:2011mc}, and black holes need not have
a spherical event-horizon topology. The general Myers-Perry metric \cite{Myers:1986un}
is rather lengthy, and the metrics for more complicated black objects (such as black rings)
are very complex. For this reason, we restrict our attention to Myers-Perry black
holes with a spherical event-horizon topology and a single axis of rotation.
In this case the general Myers-Perry metric simplifies to \cite{Myers:1986un}:
\begin{eqnarray}
ds^{2} & = &
-\left( 1- \frac {\mu }{\Sigma r^{n-1}} \right) dt^{2}
-\frac {2a\mu \sin ^{2} \theta }{\Sigma r^{n-1}} \, dt \, d\varphi
+\frac {\Sigma }{\Delta } \, dr^{2}
+ \Sigma \, d\theta ^{2}
\nonumber \\ & &
+ \left( r^{2} + a^{2} + \frac {a^{2}\mu \sin ^{2} \theta}{\Sigma r^{n-1}}
\right) \sin ^{2} \theta \, d\varphi ^{2}
+r^{2} \cos ^{2}\theta \, d\Omega _{n}^{2},
\label{eq:MyersPerry}
\end{eqnarray}
where
\begin{equation}
\Delta = r^{2} +a^{2} - \frac {\mu }{r^{n-1}}, \qquad
\Sigma = r^{2}+a^{2} \cos ^{2} \theta ,
\label{eq:MPDelta}
\end{equation}
and $d\Omega _{n}^{2}$ is the metric on the $n$-dimensional unit sphere.
The mass $M$ and angular momentum $J$ of the black hole are given by:
\begin{equation}
M=\frac {1}{16\pi }\left( n+2 \right) \mu\,A_{n+2}, \
\qquad
J= \frac {2}{n+2}\,aM,
\end{equation}
and $A_{n+2}=2 \pi^{(n+3)/2}/\Gamma[(n+3)/2]$ is the area of the
$\left( n+2 \right)$-dimensional unit sphere. The horizon radius $r_{h}$ of the black hole
is determined through the equation $\Delta(r_h)=0$ and it may be written as
$r_h^{n+1}=\mu/(1+a_*^2)$, where $a_*=a/r_h$.
When $a=0$ and the black hole is non-rotating, the metric (\ref{eq:MyersPerry})
reduces to the Schwarzschild-Tangherlini spherically-symmetric metric \cite{Tangherlini}.

In (\ref{eq:MyersPerry}), the co-ordinates $(t,r, \theta, \varphi )$ are the
co-ordinates on the brane and the $d\Omega _{n}^{2}$ is the part of the
metric coming from the extra dimensions.
To find the metric of the higher-dimensional black hole as seen by an observer
on the brane, we simply fix the co-ordinates in the extra dimensions and obtain:
\begin{eqnarray}
ds^{2} & = &
-\left( 1- \frac {\mu }{\Sigma r^{n-1}} \right) dt^{2}
-\frac {2a\mu \sin ^{2} \theta }{\Sigma r^{n-1}} \, dt \, d\varphi
+\frac {\Sigma }{\Delta } \, dr^{2}
+ \Sigma \, d\theta ^{2}
\nonumber \\ & &
+ \left( r^{2} + a^{2} + \frac {a^{2}\mu \sin ^{2} \theta}{\Sigma r^{n-1}}
\right) \sin ^{2} \theta \, d\varphi ^{2}.
\label{eq:MPbrane}
\end{eqnarray}
Note that the brane metric (\ref{eq:MPbrane}) still depends on $n$, the number of
extra dimensions. It reduces to the usual Kerr metric when $n=0$.
Although the higher-dimensional Myers-Perry metric (\ref{eq:MyersPerry}) is a
solution of the vacuum Einstein equations in $(4+n)$ dimensions,
the brane metric (\ref{eq:MPbrane}) is not a solution of the four-dimensional vacuum
Einstein equations \cite{Sampaio:2009ra}.
Instead, the space-time (\ref{eq:MPbrane}) has a non-zero classical stress-energy
tensor representing an effective fluid seen by an observer on the brane.
This arises from the fact that the black hole is a higher-dimensional object, but
a brane observer cannot directly probe the extra dimensions \cite{Sampaio:2009ra}.

\subsection{Black holes in RS brane-worlds}
\label{sec:RSbh}

In the RS brane-world, only one extra dimension is assumed to exist transverse to
our brane. The bulk is not empty but filled with a negative cosmological
constant $\Lambda_B<0$. The higher-dimensional space-time is therefore an
anti-de Sitter (AdS) space-time that contains either two (RS-I model
\cite{Randall:1999ee}) or one (RS-II model \cite{Randall:1999vf}) Minkowski
branes. The branes have a non-vanishing tension that, together
with the bulk cosmological constant, cause the warping of the metric along the
fifth dimension. In the context of the RS-I model, where two flat branes are
separated in the extra dimension by a distance a few times the Planck length,
this warping is used to address the hierarchy problem.
The RS-II model is by far the more interesting one from the gravitational
point of view: the second brane is sent to infinity, and the warping
causes the localisation of graviton close to the brane and the restoration
of four-dimensional gravity despite the presence of an infinitely-extended
extra dimension.

Substituting the Minkowski line-element on the brane by the Schwarzschild
line-element, the following brane-world solution was found soon after
in \cite{CHR}:
\begin{equation}
ds^2=e^{-2|y|/\ell_{AdS}} \left[-\left(1-\frac{2M}{r}\right)dt^2 + \left(1-\frac{2M}{r}\right)^{-1} dr^2
+ r^2\,d\Omega_2^2\right] +dy^2\,,
\label{black-string}
\end{equation}
where $\ell_{AdS}=\sqrt{-6/\Lambda_B}$ is the AdS curvature length. The
projection of the above five-dimensional solution on the brane, located at $y=0$,
has exactly the form of a four-dimensional black hole. However, it was
demonstrated that it does not describe a regular black hole localised on the
brane but rather an AdS black string that, in the context of the RS-II model,
has an infinitely-extended singularity along the bulk. In addition, it suffers
from a Gregory-Laflamme instability \cite{GL, RuthGL}.

Despite the numerous attempts to derive a regular, asymptotically AdS
black-hole solution localised on a brane with a non-vanishing tension
(for an indicative list of papers, see
\cite{tidal,Papanto,KT,KOT,CasadioNew,Frolov,Karasik,GGI,CGKM,Ovalle,KPZ}; for
a more complete list of references, see
\cite{Gregory_Aegean,Kanti:2009sz,Tanahashi}),
up to today no analytical solution has been constructed. Numerical
studies \cite{KTN, Kudoh, TT} found black-hole solutions with horizon
radius smaller than or of the order of the AdS length $\ell_{AdS}$ in the
context of five- and six-dimensional warped models. The failure to find
larger, static black-hole solutions, combined with the results
following from lower-dimensional constructions of brane-world black holes
\cite{EHM_4D,EHM_4D1, AS}, led to arguments for the non-existence of such black holes
in the context of the RS model \cite{Bruni, Dadhich, Kofinas, Tanaka, EFK, EGK}.

A central role in this conjecture is played by the AdS/CFT
correspondence: when applied in the context of the RS model \cite{HHR}, it
dictates that classical gravity in the AdS bulk is equivalent to a
strongly-coupled quantum Conformal Field Theory (CFT) living on the brane.
If a five-dimensional classical solution
exists, describing a regular black hole localised on the brane, then the
large number of CFT modes that couple to four-dimensional gravity on the brane
will cause the rapid evaporation of the black hole (we will return to this topic
in section \ref{sec:BBHrad}). Therefore, the projection of the metric on the
brane ought to describe a quantum-corrected, non-static black hole; its
classical counterpart in the bulk will then have to be non-static, too. This
argument applies only for large-mass black holes for which the quantum
corrections in the AdS bulk are negligible so that the five-dimensional solution
can be considered as classical.

There are several results in the literature  supporting the validity of
the AdS/CFT correspondence in the RS model \cite{Gregory_Aegean,Tanahashi},
such as: the agreement in the form of the Newtonian
potential on the brane, calculated through the Kaluza-Klein graviton states
or the CFT brane modes, and the automatic appearance of a radiation term, that
may be associated to the emission of brane CFT modes by the black hole,
in the Friedmann equation on the brane. But there are also counter-arguments
to the above \cite{Fitzpatrick, Zegers, Kaloper} according to which one
should not expect important quantum corrections on the brane. In support
of the latter view, recent numerical studies \cite{FW, Page, Page1} find solutions
that describe both small and large black holes in the context of the RS
model (see also \cite{Heydari, Dai_Stojkovic, Yoshino, Kleihaus, Cuadros}).

The complexity of the bulk equations and junction conditions that one should
solve to find a complete bulk/brane solution, and the non-trivial topology
of the AdS space-time background are two decisive factors contributing to
the difficulty in finding viable black-hole solutions in the RS model.
For large-mass black holes, there might be additional, more
subtle, reasons: in \cite{CGKM} it was shown that the brane trajectories
in the background of a bulk Schwarzschild-AdS black hole are more finely-tuned
for large-mass black holes; also, the recoil effect \cite{Frolov_recoil}, that
may be caused by the asymmetric emission of bulk modes resulting in the
black hole leaving the brane, is more effective for large black holes
 than for small ones \cite{Gregory_Aegean}.


\section{Hawking radiation from black holes in the ADD model}
\label{sec:MPrad}

In this section we consider Hawking radiation from black holes in the
ADD model, both on the brane and in the bulk. The presentation of the
formalism will be based on the simply-rotating Myers-Perry black hole
discussed in section \ref{sec:ADDbh} -- when necessary, the
spherically-symmetric limit may be recovered by setting $a=0$.
We first bring together all the relevant field equations for the different
types of radiation before discussing a selection of results.
The formalism for the different types of quantum field is quite involved,
and here we are attempting a unified presentation.
Some compromises in notation are inevitable in this situation.
In particular, we always label our field modes by an index $\Lambda $, although
the exact form of $\Lambda $ will vary depending on the spin of the field and
whether we are considering brane or bulk emission.

\subsection{Formalism for field perturbations}
\label{sec:formalism}

In this subsection we consider only massless particles.  The formalism outlined can be readily extended to include
mass and charge. We will briefly discuss some of the effects of mass and charge on the Hawking radiation in section \ref{sec:additional}.

\subsubsection{Teukolsky formalism on the brane}
\label{sec:branefields}

We now consider the formalism for massless particles of spin $0$, $1/2$ and
$1$ on the brane metric (\ref{eq:MPbrane}).
Teukolsky \cite{Teukolsky:1972my,Teukolsky:1973ha} developed a unified formalism for
describing perturbations of a four-dimensional Kerr black hole with these spins
(see also \cite{Chandrasekharbook}). Teukolsky's original formalism also applies to
spin-$2$ perturbations but we shall consider those separately.
Teukolsky's formalism extends easily to perturbations of spin $0$, $1/2$
and $1$ on the brane metric (\ref{eq:MPbrane}).

The Newman-Penrose formalism \cite{Newman:1961qr} is used to write the perturbation
equations for each type of particle as a single master equation for a quantity
$\Psi _{s}=\Psi _{s}(t,r,\theta ,\varphi)$.
The form of $\Psi _{s}$ depends on the spin $s$ of the field under consideration -
details can be found in \cite{Kanti:2004nr}.
The resulting Teukolsky equation for the variable $\Psi _{s}$ takes the form
\cite{Casals:2006xp}
\begin{eqnarray}
{\mathcal {J}}_{s} & = &
\left[ \frac {(r^2+a^2)^2}{\Delta}
-a^2\sin^2\theta\right]\partial _{t}{\Psi _{s}}+
\frac{2a\mu}{\Delta r^{n-1}}\partial _{t}\partial _{\varphi }{\Psi _{s}}
+
\left[\frac{a^2}{\Delta}-\frac{1}{\sin^2\theta}\right]
\partial _{\varphi }{\Psi_{s}}
\nonumber \\ & &
-\Delta^{-s}\frac{\partial}{\partial r}
\left(\Delta^{s+1}\partial _{r}{\Psi_{s}}\right)-
\frac{1}{\sin\theta}
\frac{\partial}{\partial \theta}\left(\sin\theta\partial _{\theta }{\Psi_{s}}
\right)-
2s\left[\frac{a\Delta'}{2\Delta}+\frac{i\cos\theta}{\sin^2\theta}\right]
\partial _{\varphi }{\Psi_{s}}
\nonumber \\ & &
+
2s\left[r+\bar\rho -\frac{(r^2+a^2)\Delta'}{2\Delta}\right]
\partial _{t}{\Psi_{s}}
+s\left[s\cot^2\theta-1+(2-\Delta'')\delta_{s,|s|}\right]\Psi_{s}\,,
\nonumber \\
\label{eq:TeukEq}
\end{eqnarray}
where ${\bar {\rho }}= r + i a \cos \theta$ and ${\mathcal {J}}_{s}$ is a source term,
whose details for each spin $s$ can be found in \cite{Kanti:2004nr}.
The metric function $\Delta $ is given by (\ref{eq:MPDelta}).
The Teukolsky equation (\ref{eq:TeukEq}) is separable. We write
\begin{equation}
\Psi _{s} = e^{-i\omega t} e^{im\varphi }R_{\Lambda }(r)\,S_{\Lambda }(\theta ),
\label{eq:branemode}
\end{equation}
where $\Lambda = \{s, \omega, \ell ,m \}$,
$\omega $ is the field mode frequency, $m=-\ell, -\ell + 1, \ldots , \ell -1 , \ell $
is the azimuthal quantum number and $\ell \ge s$ is the total angular momentum
quantum number. Then the following radial and angular equations are obtained
\cite{Kanti:2004nr,Casals:2006xp}:
\begin{eqnarray}
0 & = &
\Delta ^{-s} \frac {d}{dr} \left( \Delta ^{s+1} \frac {dR_{\Lambda }}{dr} \right)
+ \left[ \Delta ^{-1} \left( K_{\omega m}^{2}-isK_{\omega m} \Delta '
\right) +4is\omega r
\right. \nonumber \\  & & \left.
+s \delta _{s,\left| s \right| } \left( \Delta '' -2 \right)
-a^{2} \omega ^{2} +2ma\omega - \lambda _{\Lambda } \right] R_{\Lambda }(r) ,
\label{eq:radialbrane}
\\
0 & = &
\frac {1}{\sin \theta } \frac {d}{d\theta } \left( \sin \theta
\frac {dS_{\Lambda }}{d\theta } \right)
+\left[
-\frac {2ms\cot \theta }{\sin \theta } -\frac {m^{2}}{\sin ^{2}\theta }
+a^{2} \omega ^{2} \cos ^{2} \theta -2as\omega \cos \theta
\right. \nonumber \\ & & \left.
+s
-s^{2} \cot^{2} \theta +\lambda _{\Lambda } \right] S_{\Lambda }(\theta ) ,
\label{eq:angularbrane}
\end{eqnarray}
where
\begin{equation}
K_{\omega m}  =  \left( r^{2} +a^{2}\right) \omega - am.
\end{equation}
The angular functions $S_{\Lambda }(\theta )$ are spin-weighted spheroidal harmonics
\cite{Fackerell,Seidel:1988ue}, and they and the eigenvalues $\lambda _{\Lambda }$
have to be computed numerically when $a\omega \neq 0$.
For $a\omega =0$, the eigenvalues are:
\begin{equation}
\lambda _{\Lambda } = \ell \left( \ell +1 \right) - s\left( s+ 1\right) ,
\end{equation}
and the angular functions $S_{\Lambda }(\theta )$ reduce to spin-weighted spherical
harmonics \cite{Goldberg:1966uu}.

\subsubsection{Bulk fields}
\label{sec:bulkfields}

We now consider the equations satisfied by scalar and graviton perturbations
of the higher-dimensional bulk metric (\ref{eq:MyersPerry}).

Firstly, consider a massless scalar field propagating on the metric (\ref{eq:MyersPerry}).
The massless scalar wave equation is separable. We write the scalar field $\Psi _{0}$ as
\begin{equation}
\Psi _{0} = e^{-i\omega t} e^{im\varphi }\,R_{\Lambda }(r)\,S_{\Lambda }(\theta )
\,Y_{j n}(\Omega)\,,
\label{eq:scalarbulk}
\end{equation}
where the index $\Lambda $ is now $\{ \omega, \ell, m, j, n\}$
and $Y_{jn}(\Omega)$ is a hyper-spherical harmonic \cite{Mullerbook} depending on the
higher-dimensional bulk co-ordinates and indexed by an integer $j$.
The following radial and angular equations are obtained from the scalar field equation
\cite{Casals:2008pq}:
\begin{eqnarray}
0 & = &
\frac {1}{r^{n}} \frac {d}{dr} \left( r^{n}\Delta \frac {dR_{\Lambda }}{dr} \right)
+\left[
\Delta ^{-1} K_{\omega m}^{2}- a^{2}r^{-2}j \left( j + n -1\right)
\right. \nonumber \\ & &  \left.
-a^{2} \omega ^{2} +2ma\omega - \lambda _{\Lambda }
\right] R_{\Lambda }(r) ,
\label{eq:radialscalarbulk}
\\
0 & = &
\frac {1}{\sin \theta \cos ^{n}\theta }
\frac {d}{d\theta }
\left( \sin \theta \cos ^{n} \theta
\frac {dS_{\Lambda }}{d\theta } \right)
+\left[ \omega ^{2} a^{2} \cos^{2} \theta - \frac {m^{2}}{\sin ^{2}\theta }
-\frac {j \left(j  +n-1 \right)}{\cos ^{2}\theta }
\right. \nonumber \\  & & \left.
+\lambda _{\Lambda } \right]
S_{\Lambda }(\theta ) .
\label{eq:angularscalarbulk}
\end{eqnarray}

Gravitational perturbations are much more difficult to deal with
as Teukolsky's four-dimensional formalism does not readily extend to higher
dimensions.
A complete analysis is currently available only for higher-dimensional spherically-symmetric black holes. For spherically-symmetric black holes, a general gravitational
perturbation decomposes into three parts: a symmetric traceless
tensor $T$, a vector $V$ and a scalar part $S$ \cite{Kodama:2003jz}.
The master equation for each type of gravitational perturbation is separable
and the relevant field quantity is written in a form similar to (\ref{eq:scalarbulk})
(see \cite{Kodama:2003jz} for details).
For each type of gravitational perturbation, the radial functions satisfy the equation
\cite{Ishibashi:2003ap}
\begin{equation}
0=\left[ 1 - \left(\frac {r_{h}}{r} \right) ^{n+1} \right]
 \frac {d}{dr} \left\{ \left[ 1 - \left(\frac {r_{h}}{r} \right) ^{n+1} \right]
 \frac {dR_{\Lambda }}{dr} \right\}
+ \left[ \omega ^{2} - {\mathcal {V}}_{\Lambda } \right] R_{\Lambda }(r) ,
\label{eq:gravbulk}
\end{equation}
where the form of the potential ${\mathcal {V}}_{\Lambda }$
depends on the type of gravitational perturbation.
The angular functions are simply spin-weighted hyper-spherical harmonics.
The index $\Lambda $ now takes the form $\{ B, \omega, \ell ,n \}$ where
$B\in \{ S, V, T\}$
indicates whether we are considering a scalar ($S$), vector ($V$) or tensor ($T$)
type of gravitational perturbation and the other labels are as before.
For tensor-like and vector-like perturbations the
potential ${\mathcal {V}}_{\Lambda }$ is \cite{Ishibashi:2003ap}
\begin{equation}
{\mathcal {V}}_{T/V, \omega ,\ell ,n }=
\frac {1}{r^{2}} \left[ 1 - \left(\frac {r_{h}}{r} \right) ^{n+1} \right]
\left[
\ell \left( \ell  +n  + 1\right) + \frac {n\left( n + 2\right)}{4}
-\frac {k}{4} \left( n+2 \right) ^{2}
\frac {r_{h}^{n+1}}{r^{n+1}}
\right],
\end{equation}
where $k=-1$ for tensor-like ($T$) perturbations and $k=3$ for vector-like ($V$)
perturbations.
For scalar-like ($S$) graviton perturbations, the potential has the more complicated
form \cite{Ishibashi:2003ap}:
\begin{equation}
{\mathcal {V}}_{S,\omega, \ell , n} = \frac {1}{r^{2}}
\left[ 1 - \left(\frac {r_{h}}{r} \right) ^{n+1} \right]
\frac {qx^{3}+px^{2}+wx+z}{4\left[
2u + (n+2)(n+3)x
\right] ^{2} }\,,
\end{equation}
where
\begin{equation}
x=\frac {r_{h}^{n+1}}{r^{n+1}}, \qquad
u = \ell \left( \ell +n+1 \right) -n-2 ,
\end{equation}
and
\begin{eqnarray}
q & = & \left( n+2 \right)^{4} \left( n+3 \right) ^{2},
\nonumber \\
p & = & \left( n+2 \right) \left( n+3 \right)\left[
4u\left( 2n^{2}+5n+6\right)
+n\left( n+2 \right) \left( n+3\right) \left( n-2\right)
\right] ,
\nonumber \\
w & = & -12u\left(n+2 \right) \left[
u\left( n-2 \right) +n \left( n + 2 \right) \left( n+3 \right)
\right] ,
\nonumber \\
z & = &
16u^{3}+4u^{2} \left( n+2 \right) \left( n+4 \right) .
\end{eqnarray}

For rotating higher-dimensional black holes, the general gravitational perturbation
equations are much more complicated \cite{Durkee:2010ea,Reall:2012it,Murata:2011zz}.
In general they are not separable, which means that a computation of the
Hawking radiation for gravitons has to date proved intractable.
However, some progress can be made in the case where the higher-dimensional
gravitational background is the warped product of an $m$-dimensional
space-time ${\cal N}$ and an $n$-dimensional space ${\cal K}$ of constant
curvature, a class of backgrounds that includes the simply-rotating Myers-Perry
metric (\ref{eq:MyersPerry}). In that case, the equations for tensor-type perturbations
simplify considerably \cite{Kodama:2007sf,Kodama:2007ph} and are separable.
The radial and angular equations take the forms
(\ref{eq:radialscalarbulk}--\ref{eq:angularscalarbulk}) respectively, that is,
the equations for the scalar field modes in the bulk -- the only difference is that
$\ell \ge 0$ for scalars and $\ell \ge 2$ for gravitons \cite{Kodama:2007ph}.

\subsection{Grey-body factors and fluxes}
\label{sec:greybody}

We are interested in the fluxes of particles $N$, energy $E$ and angular momentum
$J$ for the various different fields.
The differential fluxes per unit time and
unit frequency $\omega $ take the form:
\begin{eqnarray}
\frac {d^{2}N}{dt\, d\omega} & = &
\frac {1}{2\pi }
\sum _{j}
\sum _{\ell = s}^{\infty } \sum _{m=-\ell }^{\ell }
\frac {1}{e^{{\tilde {\omega }}/T_{H}}\pm 1}
\,{\mathcal {N}}_{\Lambda } {\Gamma }_{\Lambda },
\label{eq:particleflux}
\\
\frac {d^{2}E}{dt\, d\omega } & = &
\frac {1}{2\pi }
\sum _{j}
\sum _{\ell = s}^{\infty } \sum _{m=-\ell }^{\ell }
\frac {\omega }{e^{{\tilde {\omega }}/T_{H}}\pm 1}
\,{\mathcal {N}}_{\Lambda } {\Gamma }_{\Lambda },
\label{eq:powerflux}
\\
\frac {d^{2}J}{dt \, d\omega } & = &
\frac {1}{2\pi }
\sum _{j}
\sum _{\ell = s}^{\infty } \sum _{m=-\ell }^{\ell }
\frac {m}{e^{{\tilde {\omega }}/T_{H}}\pm 1}
\,{\mathcal {N}}_{\Lambda } {\Gamma }_{\Lambda }.
\label{eq:angmomflux}
\end{eqnarray}
Here we have written out precisely the mode sums represented schematically in
(\ref{eq:simpleparticleflux}, \ref{eq:simplefluxes1}).
As well as the usual sums over the angular momentum quantum numbers $\ell ,m$,
there is an additional sum over $j$ for scalar field emission in the bulk and
tensor-type graviton emission from a rotating black hole, where $j$ indexes the
hyper-spherical harmonics in these cases.
There is no sum over $j$ for graviton emission from spherically-symmetric
higher-dimensional black holes.
In the thermal factor, the $+$ sign is for fermionic fields and the $-$ sign
for bosonic fields. In the above, ${\tilde {\omega }}$ is given by (\ref{eq:tildeomega}),
while the temperature $T_H$ and angular velocity $\Omega_H$ of
the simply-rotating Myers-Perry black hole (\ref{eq:MyersPerry}) are found to be:
\beq
T_{H}=\frac{(n+1)+(n-1)a_*^2}{4\pi(1+a_*^2)r_{h}}\,,
\qquad \quad
\Omega_H= \frac{a}{r_h^2 +a^2}\,.
\label{T-Omega}\eeq

For each mode, the fluxes (\ref{eq:particleflux}--\ref{eq:angmomflux}) depend on the
grey-body factor $\Gamma _{\Lambda }$, and also a degeneracy factor
${\mathcal {N}}_{\Lambda }$ accounting for the multiplicity of modes having the
quantum numbers $\{ \omega ,\ell ,m, j \}$.
The degeneracy factors are always independent of the mode frequency $\omega $
and azimuthal quantum number $m$, but depend on $\ell $, $j$ (where applicable) and
the number of extra dimensions $n$.

On the brane, for fields of
spin-$1/2$ and spin-$1$, there are field modes with two polarizations,
so to take this into account we set the degeneracy factors equal to
\begin{equation}
{\mathcal {N}}_{\Lambda }
= \left\{
\begin{array}{ll}
1 & {\mbox { for $s=0$,}}
\\
2 & {\mbox { for $s=\frac {1}{2}$,}}
\\
2 & {\mbox { for $s=1$.}}
\end{array}
\right.
\label{eq:branedegeneracy}
\end{equation}
For bulk scalar fields, the degeneracy factor is \cite{Casals:2008pq}:
\begin{equation}
{\mathcal {N}}_{\Lambda }
=\frac {\left( 2j+n-1 \right) \left( j+n-2 \right) !}{j! \left(n-1 \right) !}.
\label{eq:scalarbulkdegeneracy}
\end{equation}
For bulk graviton fields, we need to consider each type of gravitational perturbation
separately. If we consider a rotating black hole, we only have separable field equations
for tensor-type gravitational perturbations, in which case the degeneracy
factor is \cite{Kanti:2009sn}
\begin{equation}
{\mathcal {N}}_{\Lambda }
=
\frac {\left( n + 1 \right) \left( n -2 \right) \left( n + j\right)
\left( j -1 \right) \left( n + 2j -1 \right) \left( n + j -3 \right)!}{
2\left( j + 1 \right) ! \left( n-1 \right) !}.
\label{eq:gravdegen2}
\end{equation}
If, on the other hand, we consider a non-rotating black hole, all three types of
gravitational perturbation (scalar $S$, vector $V$ and tensor $T$) can be considered.
The degeneracy values are then \cite{Creek:2006ia,Rubin1,Rubin2}:
\begin{eqnarray}
{\mathcal {N}}_{S, \omega, \ell, n} & = &
\frac {\left( 2 \ell + n + 1 \right) \left( \ell + n\right) !}{\left( 2\ell + 1 \right)
\ell ! \left( n + 1 \right) !},
\nonumber \\
{\mathcal {N}}_{V,\omega ,\ell ,n} & = &
\frac {\ell \left( \ell + n + 1 \right) \left( 2\ell + n + 1 \right)
\left( \ell + n -1 \right) !}{\left( 2 \ell + 1 \right)
\left( \ell + 1\right) ! n!},
\nonumber \\
{\mathcal {N}}_{T, \omega , \ell ,n} & = &
\frac {n\left( n + 3 \right) \left( \ell + n + 2 \right)
\left( \ell - 1\right)  \left( 2\ell + n + 1 \right) \left( \ell + n -1 \right) !}{
2\left( 2\ell + 1 \right) \left( \ell + 1 \right) !\left( n+1 \right) !}.
\label{eq:gravdegen1}
\end{eqnarray}

To compute the Hawking fluxes (\ref{eq:particleflux}--\ref{eq:angmomflux}), it remains
to find the grey-body factors $\Gamma _{\Lambda }$.
These are computed by numerically integrating the relevant radial equation
(\ref{eq:radialbrane}, \ref{eq:radialscalarbulk}, \ref{eq:gravbulk}).
For an ``up'' mode, the
grey-body factor $\Gamma _{\Lambda }$ is the ratio
of the flux in the mode at infinity
and the flux in the out-going part of the mode near the event horizon,
in other words it is the transmission coefficient for each ``up'' mode.
The exact form of the flux depends on the spin of the field considered
(see \cite{Kanti:2004nr,Chandrasekharbook} for details).
Here we simply state the results for the grey-body factors in each case.

We first consider scalar and graviton fields, which are each described by
a single radial function $R_{\Lambda }$ which satisfies the relevant
radial equation: (\ref{eq:radialbrane}) for scalar fields on the brane;
(\ref{eq:radialscalarbulk}) for scalar fields in the bulk;
(\ref{eq:gravbulk}) for
all types of graviton emission from a non-rotating black hole; or
(\ref{eq:radialscalarbulk}) for tensor-type graviton emission from a simply-rotating
black hole.
The ``up'' modes then have radial functions of the form:
\begin{equation}
R_{\Lambda } \sim \left\{
\begin{array}{ll}
\left( r-r_{h} \right) ^{i{\tilde {\omega }}/4\pi T_{H}}
+ C_{R,\Lambda } \left( r-r_{h} \right) ^{-i{\tilde {\omega }}/4\pi T_{H}}
& \qquad r\rightarrow r_{h}
\\
C_{T,\Lambda } r^{-y}e^{i\omega r} &
\qquad r\rightarrow \infty ,
\end{array}
\right.
\label{eq:wave}
\end{equation}
where $C_{R,\Lambda }$ and $C_{T,\Lambda }$ are complex constants, and
\begin{equation}
y = \left\{
\begin{array}{ll}
1 & {\mbox { for brane emission of scalars,}}
\\
1+\frac {n}{2} & {\mbox { for bulk emission of scalars, and tensor-type
graviton emission}}
\\
& {\mbox { from rotating black holes,}}
\\
0  & {\mbox { for graviton emission from non-rotating black holes.}}
\end{array}
\right.
\end{equation}
The grey-body factor is then simply
\begin{equation}
{\Gamma }_{\Lambda } = 1-\left| C_{R,\Lambda } \right| ^{2}=
\frac {\omega }{{\tilde {\omega }}} \left| C_{T,\Lambda } \right| ^{2}.
\label{eq:greybody}
\end{equation}
If the black hole is non-rotating, ${\tilde {\omega }}=\omega $ and
$\Gamma _{\Lambda } = | C_{T,\Lambda }| ^{2}$.
If the black hole is rotating, for modes with $\omega /{\tilde {\omega }}<0$,
equation (\ref{eq:greybody}) implies that $\Gamma _{\Lambda } <0$,
so we have super-radiance
\cite{Chandrasekharbook}.

We next consider fermion (spin-$\frac{1}{2}$) and gauge boson (spin-$1$) fields, for
which there are two radial functions, corresponding to $s=+\left| s \right| $
and $s=-\left| s \right| $.
The radial functions $R_{\Lambda }$ satisfy (\ref{eq:radialbrane}) and, for an ``up''
mode, have the asymptotic forms \cite{Kanti:2004nr,Casals:2006xp}
\begin{eqnarray}
R_{\Lambda }^{s=+\left| s\right| } & \sim &
\left\{
\begin{array}{ll}
C_{R,\Lambda }\Delta ^{-s} \left( r-r_{h} \right) ^{-i{\tilde {\omega }}/4\pi T_{H}}
& \qquad r\rightarrow r_{h}
\\
0 &
\qquad r\rightarrow \infty ,
\end{array}
\right.
\nonumber \\
R_{\Lambda }^{s=-\left| s\right| } & \sim &
\left\{
\begin{array}{ll}
\left( r-r_{h} \right) ^{i{\tilde {\omega }}/4\pi T_{H}}
& \qquad r\rightarrow r_{h}
\\
C_{T,\Lambda } r^{-\delta _{s,1}} e^{i\omega r} &
\qquad r\rightarrow \infty ,
\end{array}
\right.
\end{eqnarray}
for complex constants $C_{R,\Lambda }$ and $C_{T,\Lambda }$.
For gauge bosons with $\left| s\right| =1$, the grey-body factor is given by
(\ref{eq:greybody}), and there is super-radiance for modes with
${\tilde {\omega }}<0$.
For fermion fields with $\left| s \right| =\frac {1}{2}$,
the grey-body factor is:
\begin{equation}
\Gamma _{\Lambda } = 1-\left| C_{R,\Lambda } \right| ^{2} =
 \left| C_{T,\Lambda } \right| ^{2}.
\end{equation}
For fermions, we therefore have $\Gamma _{\Lambda }>0$ for all modes and no
super-radiance \cite{Chandrasekharbook}.


\subsection{Emission of massless fields on the brane}
\label{sec:Hawking-rad-brane}

We now present a selection of results on the decay of
higher-dimensional black holes through the emission of Hawking radiation.
The presentation of the results will be by no means exhaustive, rather
we hope that it will reveal some of the main features of the radiation
spectra from these black holes. We will start from the emission of particles
along the brane, then consider the bulk emission and finish with a
discussion of the energy balance between the two decay channels.

For a brane-localised observer, the emission of particles along the brane
is the only observable decay channel of a higher-dimensional black hole.
Drawing information from black holes in four dimensions, we expect that
higher-dimensional black holes will emit Hawking radiation during both their
rotating and spherically-symmetric phase. The rotating phase is the most
generic, however, it is also the most technically involved. Therefore, we will
start from the spherically-symmetric phase, that, although it is chronologically
second, has a significantly simpler treatment.

\subsubsection{Non-rotating black holes}

The gravitational background describing the space-time around a non-rotating,
higher-dimensional black hole, i.e. the Schwarzschild-Tangerlini line-element
\cite{Tangherlini},
and the corresponding field equations follow easily from (\ref{eq:MPbrane})
and (\ref{eq:radialbrane}--\ref{eq:angularbrane}), respectively, by setting $a=0$.
In particular, the angular equation (\ref{eq:angularbrane})  now reduces to
the one for the spin-weighted spherical harmonics with a well-defined eigenvalue,
and offers no new information given the spherically-symmetric emission. Therefore,
it is only the radial equation (\ref{eq:radialbrane}), significantly simplified
after setting $a=0$, that needs to be integrated. This has been performed both
analytically \cite{KMR1, KMR2, IOP1} and numerically \cite{HK}. In the former case,
an approximation technique needs to be applied: in this, the radial equation is
solved in the two asymptotic regimes, i.e. near the horizon ($r \simeq r_h$) and
far away from it ($r \gg r_h$), and the two solutions are matched at an
intermediate point. The analytic result derived for the grey-body factor is
valid only under the assumption that the energy of the emitted particle satisfies
the constraint $\omega r_h \ll 1$. Therefore, for the derivation of the
radiation spectra beyond the low-energy regime, one needs to employ numerical
methods.

\begin{figure}
\centering
\mbox{\includegraphics[width=0.48\textwidth]{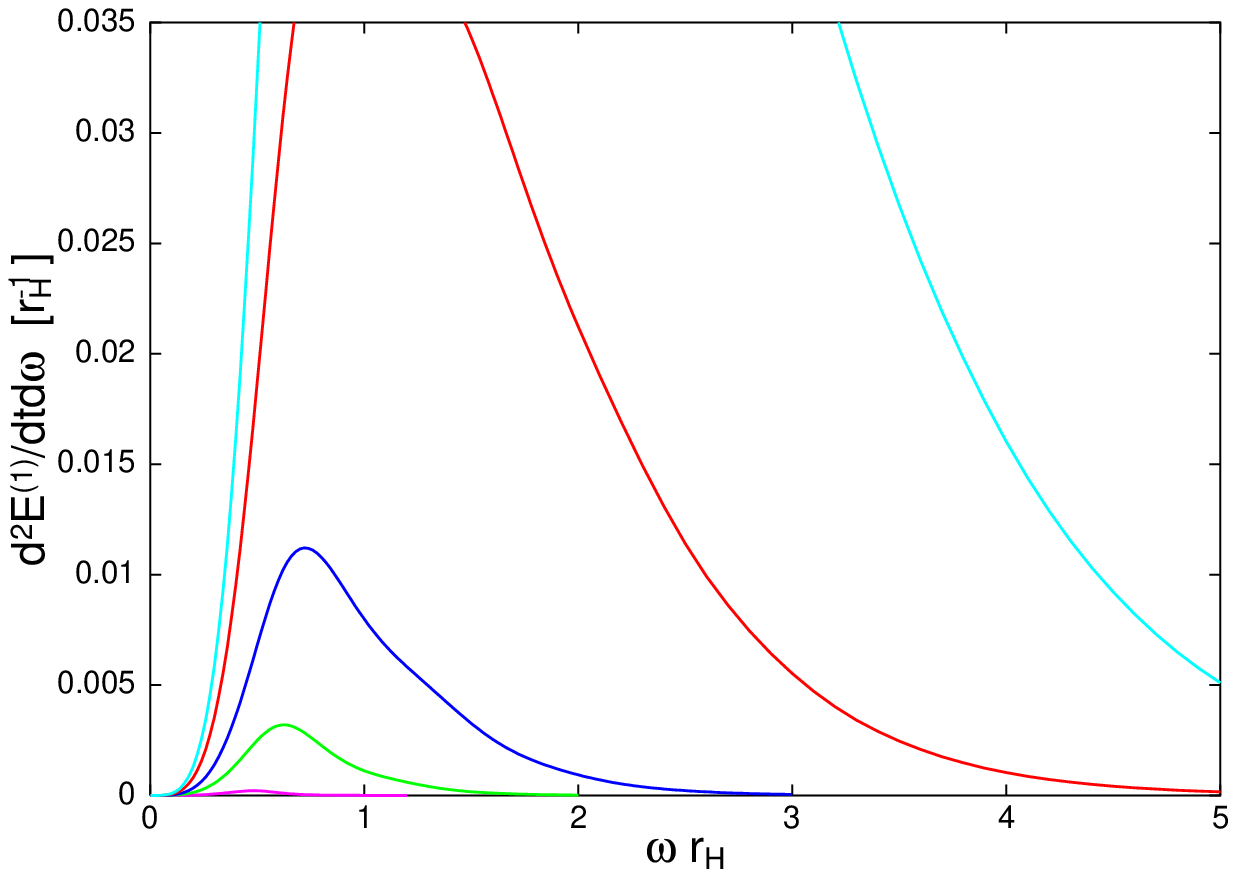}}
{\includegraphics[width=0.48\textwidth]{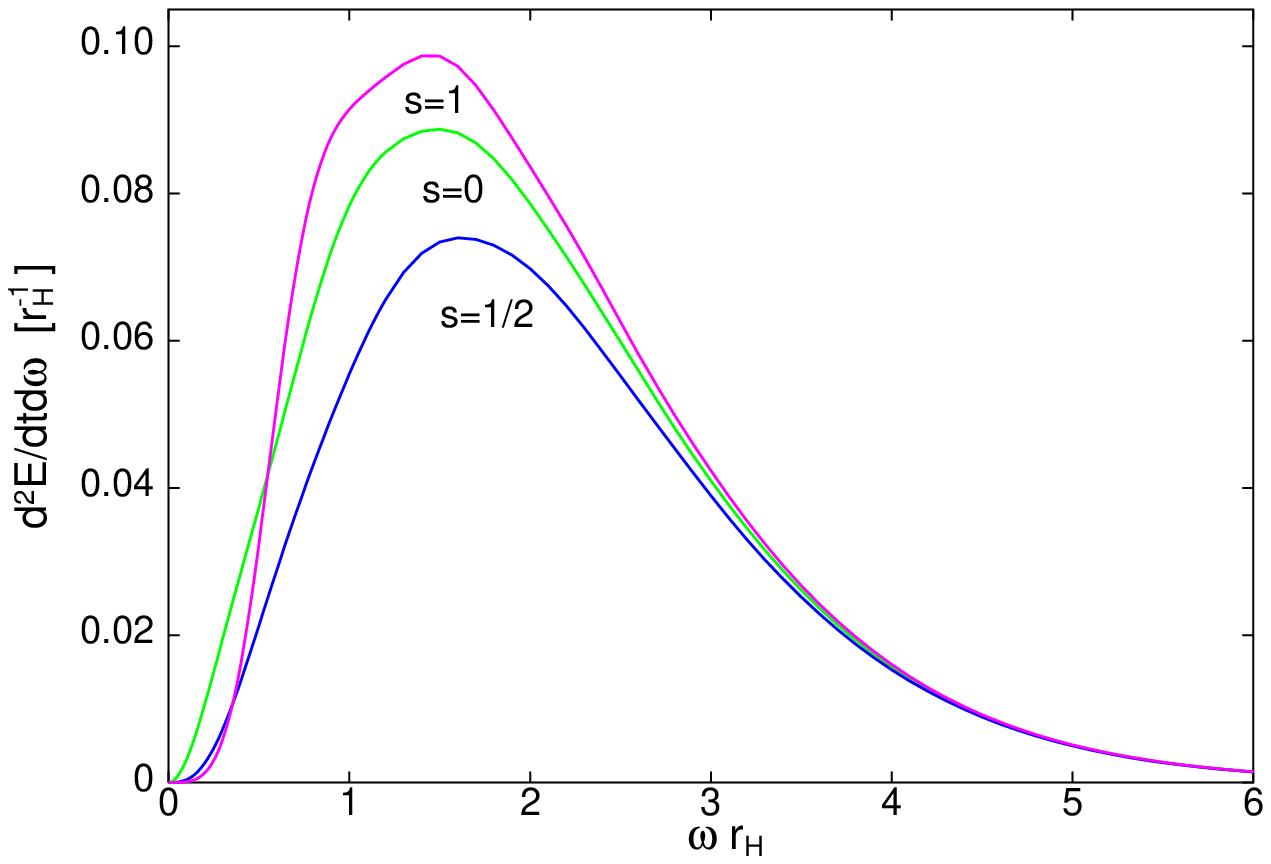}}
\caption{Energy emission rates {\bf (a)} for gauge bosons, for $n=0,1,2,4$ and 6 (from
bottom to top) \cite{Kanti:2004nr}, and {\bf (b)} for all species of brane-localised
particles for $n=6$ \cite{HK}.}
\label{brane-spherical}
\end{figure}

In \cite{HK}, the complete spectra for all types of brane particles and for arbitrary
energy $\omega$ were thus numerically derived. In Figure \ref{brane-spherical}(a),
we depict the
differential energy emission rate per unit time and unit frequency for gauge bosons
emitted on the brane by a spherically-symmetric black hole, for variable $n$
\cite{Kanti:2004nr,HK}. We observe that, as the number of extra dimensions
that are transverse to the brane increases, the black hole radiates more energy
per unit time and over a much wider spectrum of frequencies.
This is due to the combined effect of the grey-body factors $\Gamma_\Lambda$ and
the temperature $T_H$ of the black hole (\ref{T-Omega}), with the latter being
clearly an increasing function of $n$, for a fixed horizon radius $r_h$.
The same behaviour is observed for all types of brane particles,
i.e. scalars, fermions and gauge bosons \cite{Kanti:2004nr,HK}. By integrating
the energy spectra over $\omega$, we find that the total emissivities are
enhanced by a factor of the order of $10^3$-$10^4$, as $n$ increases from 0 to 7.

The exact value of the enhancement factor, however, depends on the spin $s$ of the
particle, and that leads to the question whether the black hole prefers to emit
different species of particle for different values of $n$.
In Figure \ref{brane-spherical}(b), we depict the power fluxes for particles with
spin $s=0$, $1/2$ and $1$, for the case $n=6$ \cite{HK}: the gauge bosons clearly
dominate over both scalars and fermions. This result is to be compared to
the four-dimensional one \cite{Page:1976df, Sanchez1, Sanchez2, Sanchez3} where scalars dominate,
and to the case where $n$ takes intermediate values, where the
black hole emits almost equal amounts of energy in the three particle channels
\cite{HK}. One could then propose that the emission by a higher-dimensional,
spherically-symmetric black hole on the brane could reveal the number of
additional spacelike dimensions \cite{KMR1, KMR2, HK}.

\subsubsection{Rotating black holes}

\begin{figure}[t!]
\hspace*{-0.5cm}
\mbox{\includegraphics[width=0.6\textwidth,height=0.23\textheight]{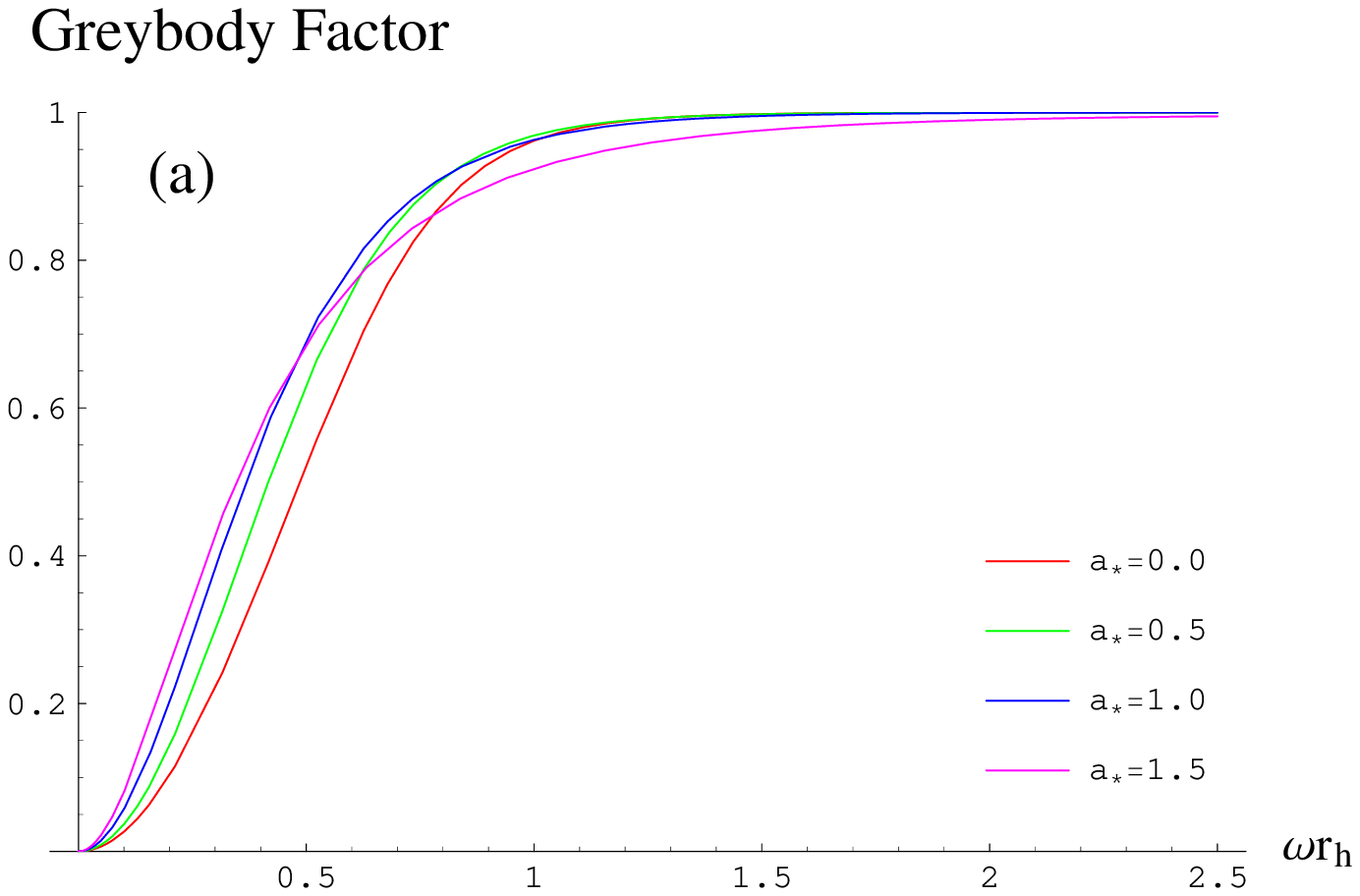}}
\hspace*{-6.3cm}
{\includegraphics[width=0.6\textwidth,height=0.23\textheight]{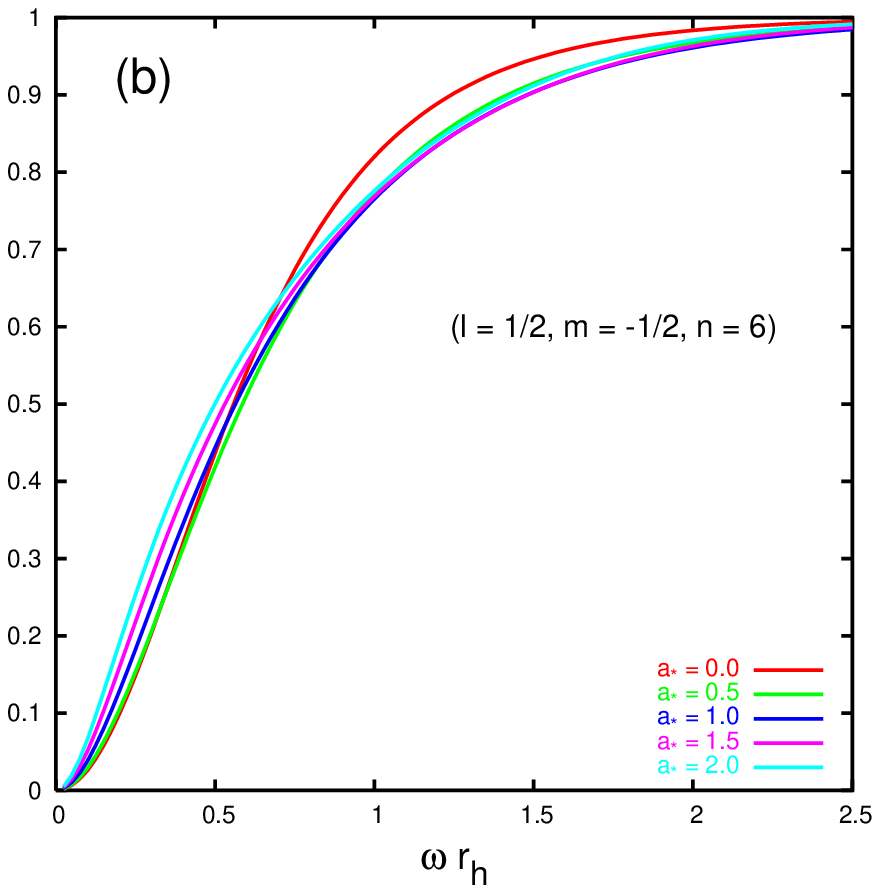}}
\caption{Grey-body factors of brane-localised fermions derived {\bf (a)}
analytically \cite{CEKT3} and {\bf (b)} numerically \cite{Casals:2006xp}
for $\ell=-m=1/2$, $n=6$ and variable $a_*$.}
\label{Grey-bosons-brane}
\end{figure}

We now turn to the preceding rotating phase that, for black holes created during a
non-head-on collision, is the most generic and perhaps the only phase realised
due to their short life-time. In this case, the space-time around the black hole
is not spherically-symmetric -- instead, the axis of rotation provides a preferred
direction in space. Therefore, the angular equation, too, contains vital information
about the emission process, and it is thus the set of equations
(\ref{eq:radialbrane}--\ref{eq:angularbrane}) that we now need to solve.
The radial equation will provide us again with the value of the grey-body factor.
For this, we need the eigenvalue $\lambda_\Lambda$ whose value may be found by
numerically integrating the angular equation. There is, however, an infinite
power-series expansion \cite{Fackerell, Seidel:1988ue, Staro}, in the limit of
small $a\omega$, of the form
\beq\lambda_{\Lambda}=-s(s+1) + \sum_k\, f_k\,(a \omega)^k=
\ell(\ell+1)-s(s+1)-\frac{2ms^2}{\ell(\ell+1)}a\omega+...\,. \label{Eigen-brane}
\eeq
The use of the above expression for $\lambda_\Lambda$ allows
for the analytical solution of the radial equation \cite{IOP1, CEKT2, CEKT3}
for the rotating phase, too, and the derivation of a formula for the grey-body
factor. As before, the validity of the result is limited: it
applies only for emission by a slowly-rotating black hole ($a_* < 1$) in the
low-energy regime ($\omega r_h \ll 1$); for the sake of comparison, in
Figure \ref{Grey-bosons-brane}(a,b), we present both the analytic \cite{CEKT3}
and the numerical result \cite{Casals:2006xp}, respectively,
for the grey-body factor of brane-localised fermions for the mode
$\ell=-m=1/2$ in the 10-dimensional case and for various values of $a_*$.
The agreement between the two results is very good in the low-energy part
of the spectrum and for small angular-momenta, however, it clearly worsens
as either $\omega$ or $a_*$ increases.

The complete radiation spectra, under no restrictions on the energy and angular
momentum, were derived in \cite{Duffy:2005ns, Casals:2005sa, Casals:2006xp, HK2}
for brane-localised scalars, fermions and gauge bosons by employing numerical
techniques. In each case, the angular equation (\ref{eq:angularbrane}) was
integrated first to derive the exact values of both the angular eigenvalue
$\lambda_\Lambda$ and the spin-weighted spheroidal harmonics $S_\Lambda$.
The radial equation (\ref{eq:radialbrane}) was solved next from the horizon
to infinity under the appropriate boundary conditions. The numerical integration
of the set of radial and angular equations presents different challenges for
different species of fields -- we address the interested reader to
\cite{Duffy:2005ns, Casals:2005sa, Casals:2006xp, HK2} for further information
on how to overcome these.

\begin{figure}[t!]
\centering
\mbox{\includegraphics[width=0.49\textwidth]{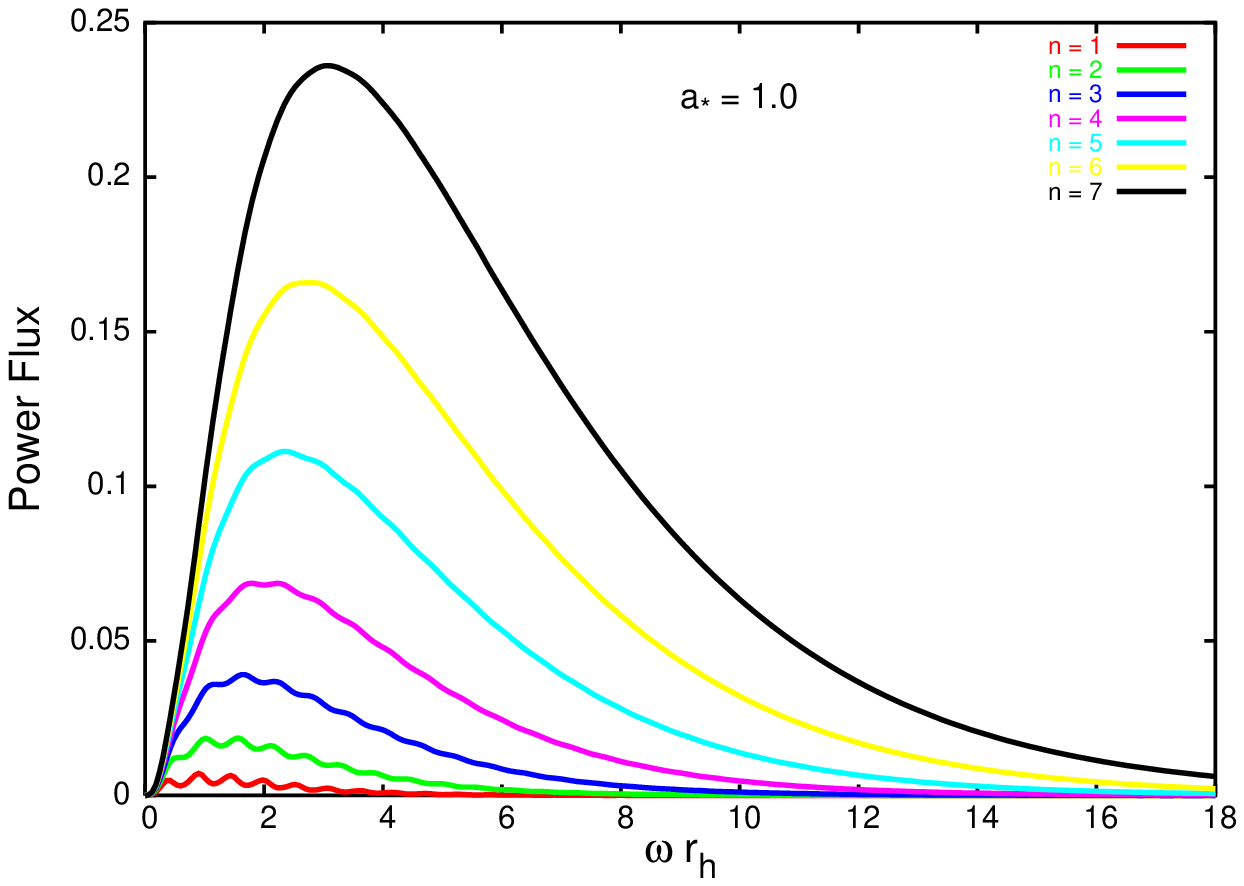}}
{\includegraphics[width=0.49\textwidth]{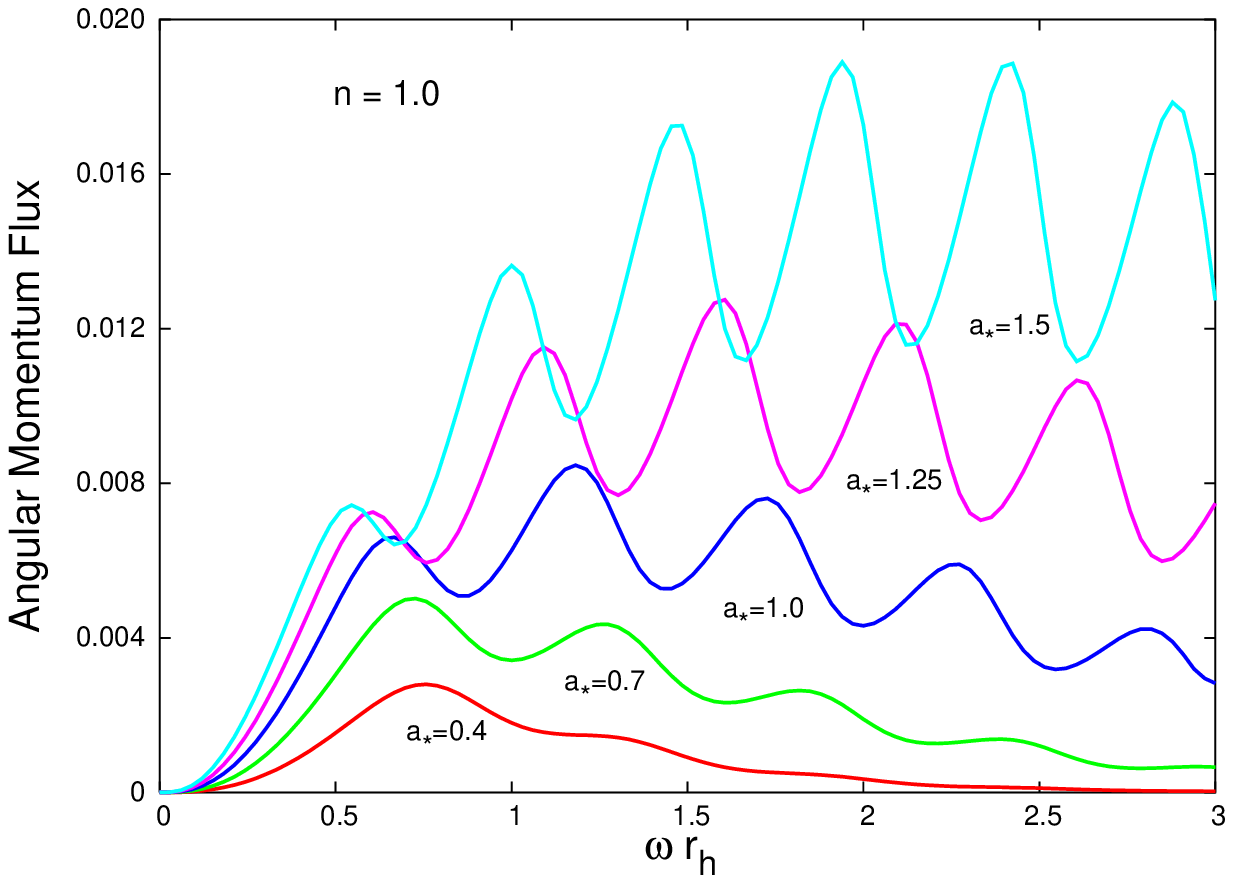}}
\caption{{\bf (a)} Energy emission rate for fermions in terms of $n$, for $a_*=1$
\cite{Casals:2006xp}, and {\bf (b)} angular-momentum emission rate
for scalars, in terms of $a_*$, for $n=1$ \cite{Duffy:2005ns}.}
\label{emission-brane}
\end{figure}

For a rotating black hole, the grey-body factor depends both on the number
of extra dimensions $n$ and the angular-momentum $a_*$ of the black hole, but
also on the part of the energy spectrum and the particular mode considered.
An indicative result for the grey-body factor for fermions \cite{Casals:2006xp}
with $\ell=-m=1/2$, $n=6$ and various values of $a_*$ was presented in
Figure \ref{Grey-bosons-brane}(b). Using the values of the numerically-derived
grey-body factors, one may proceed to determine the fluxes of particles $N$,
energy $E$ and angular momentum $J$ for a rotating black hole on the brane.
The profile of each flux shows an enhancement, in terms of both $n$ and $a_*$,
for all species of brane-localised particles.
In Figure \ref{emission-brane}(a,b), we depict indicative cases of the energy
emission rate for fermions, for $a_*=1$ and various values of $n$
\cite{Casals:2006xp}, and of the angular-momentum emission rate for scalars,
for $n=1$ and various values of $a_*$ \cite{Duffy:2005ns}, respectively.
The enhancement factor in the total emissivity of all three fluxes, when $n$
varies from 1 to 7, is typically of ${\cal O}(100)$ while the one when $a_*$
changes from 0 to 1 is of ${\cal O}(10)$.

\begin{figure}[b!]
\centering
\mbox{\includegraphics[width=0.37\textwidth]{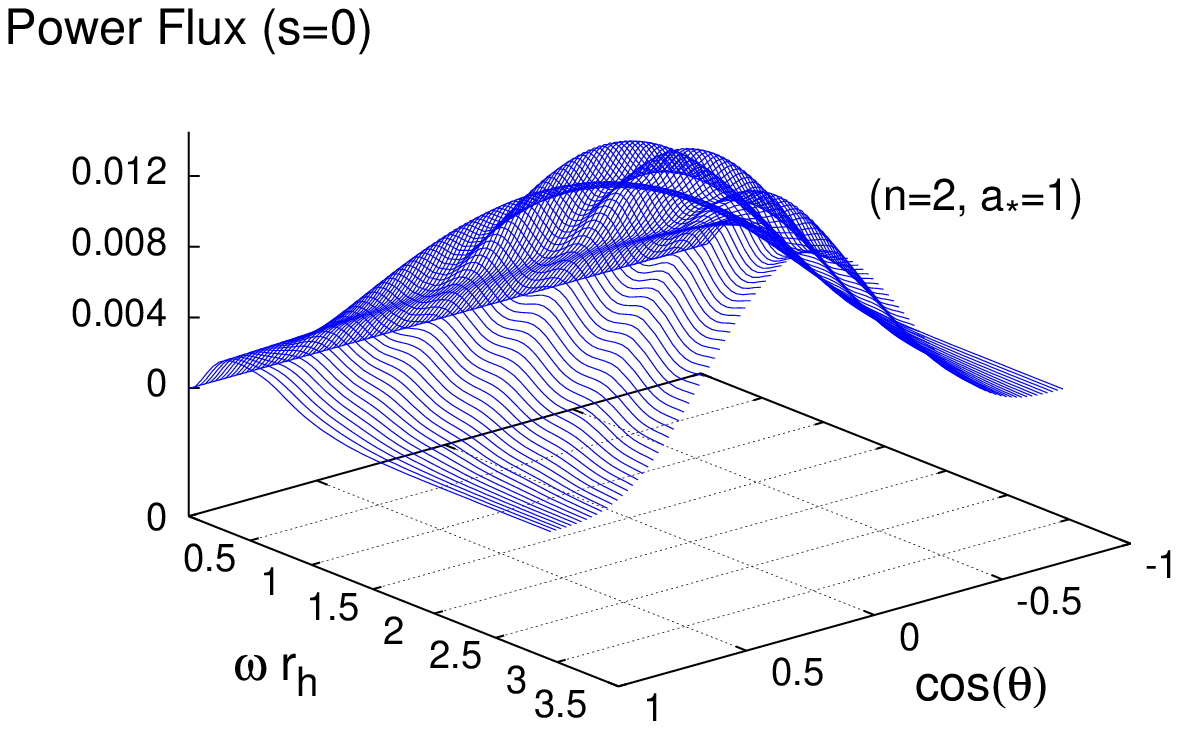}}
\hspace*{-0.85cm}
{\includegraphics[width=0.37\textwidth]{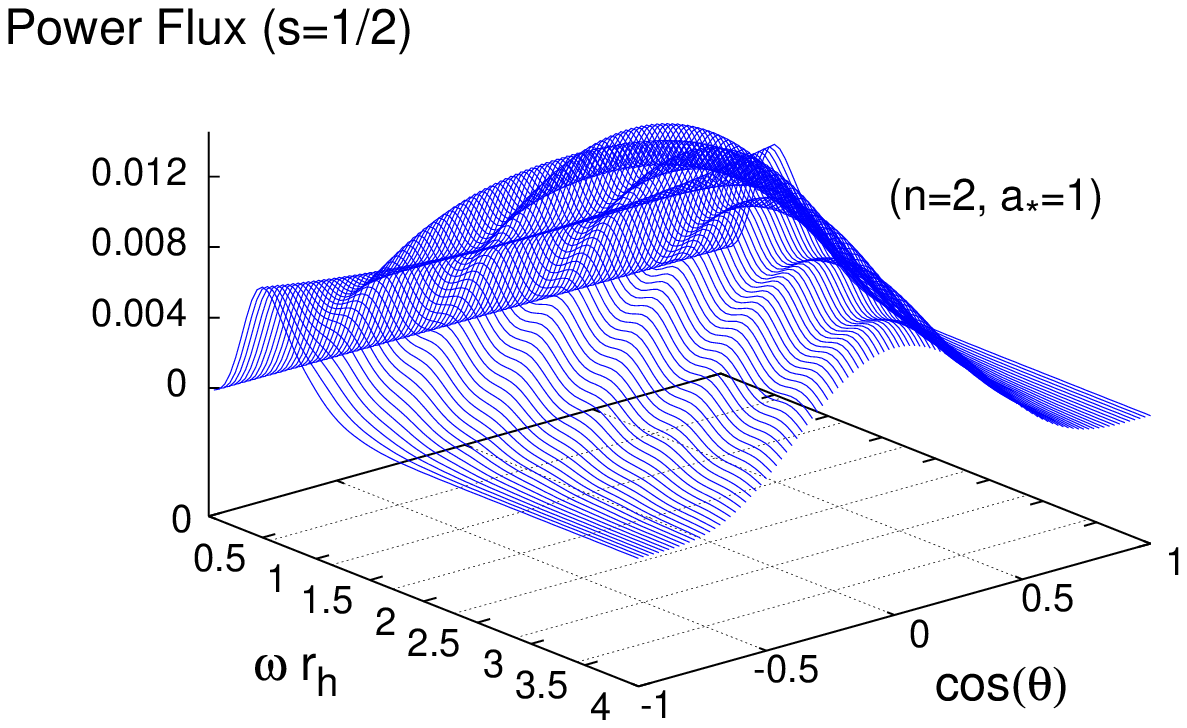}}
\hspace*{-0.85cm}
{\includegraphics[width=0.37\textwidth]{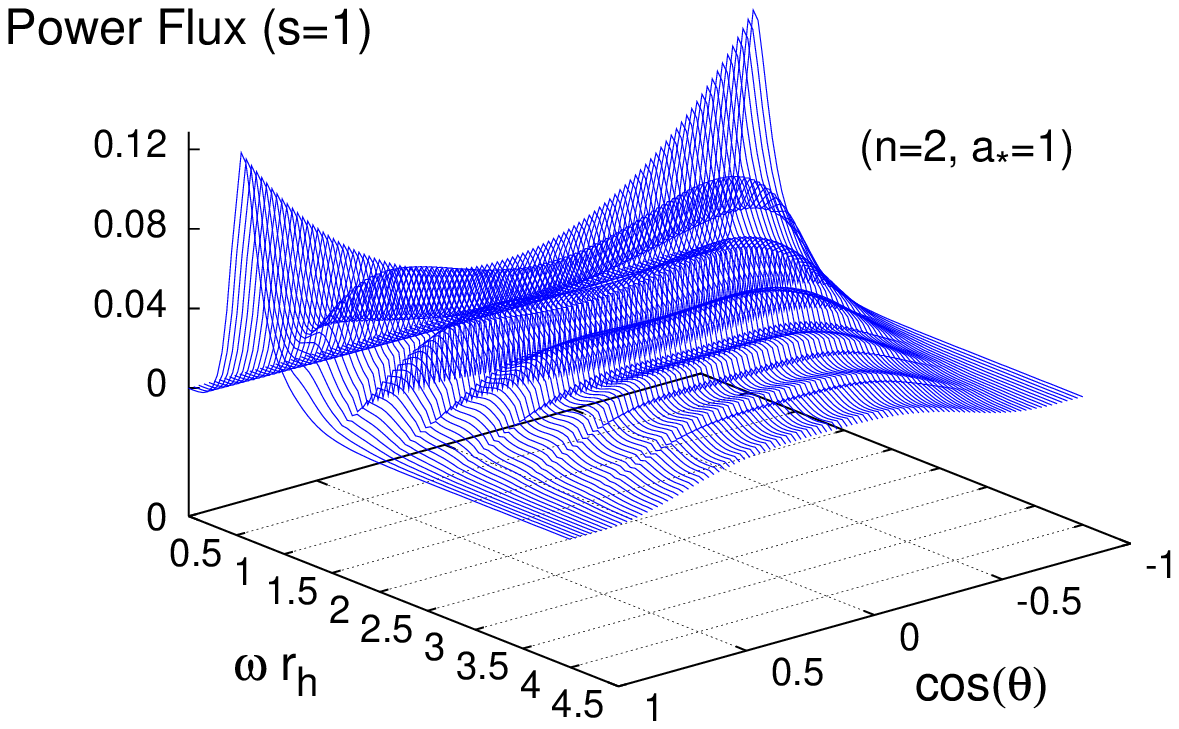}}
\caption{Angular distribution of the power spectra for scalars \cite{Duffy:2005ns}
(left plot), fermions \cite{Casals:2006xp} (middle plot) and gauge bosons
\cite{Casals:2005sa} (right plot) for $n=2$ and  $a_*=1$.}
\label{ang-distrib}
\end{figure}

The angular equation (\ref{eq:angularbrane}) is also a source of valuable
information for the emission process: the spin-weighted spheroidal harmonics
$S_\Lambda(\theta)$ contain information on the angular distribution
of the emitted particles. The differential fluxes
(\ref{eq:particleflux}--\ref{eq:angmomflux}) are derived
by integrating the appropriate operators over a
sphere at infinity. If we therefore take a step back, one may derive the
differential emission rates per unit time, unit energy and unit of $\cos\theta$,
where $\theta$ is the angle measured from the rotation axis of the black hole.
All fluxes exhibit a non-trivial angular
distribution as a result of two factors: (i) the centrifugal force, that
forces all types of particles to be emitted on the equatorial plane,
particularly for large values of $\omega$ or $a_*$, (ii) the
spin-rotation coupling, which tends to align all particles with non-vanishing
spin with the rotation
axis -- this factor has a different effect on different radiative components
and is more prominent the larger the value of the spin and the smaller the
energy of the particle. In Figure \ref{ang-distrib}, we present
three-dimensional graphs depicting the differential energy emission rate
in terms of $\omega_*$ and $\cos\theta$ for scalars, fermions and gauge
bosons, for $n=2$ and $a_*=1$ \cite{Duffy:2005ns,
Casals:2006xp, Casals:2005sa}, that clearly present the above behaviour.
Note, that for fermions and gauge bosons, the distribution is symmetric
over the two hemispheres since both radiative components, $s=\pm \frac{1}{2}$
and $s=\pm 1$, respectively, have been taken into account in the expression
of the energy emission rates in each case.

The aforementioned angular distribution of the emitted particles will be a
distinct observable effect, and will last for as long as the angular-momentum
of the black hole is non-zero. However, for a rotating black hole, the task
of drawing quantitative information from the Hawking radiation spectra,
regarding the parameters of space-time, presents a serious
difficulty: both the number of extra spacelike dimensions $n$ and the
angular-momentum parameter $a_*$ cause an enhancement of the emission rates
(\ref{eq:particleflux}--\ref{eq:angmomflux}).
One therefore needs to break this degeneracy, by using an observable that
would depend rather strongly on the value of only one of these parameters
and, at the same time, be almost insensitive to the value of the other.

\begin{figure}[b!]
\begin{center}
\mbox{
\includegraphics[scale=0.58]{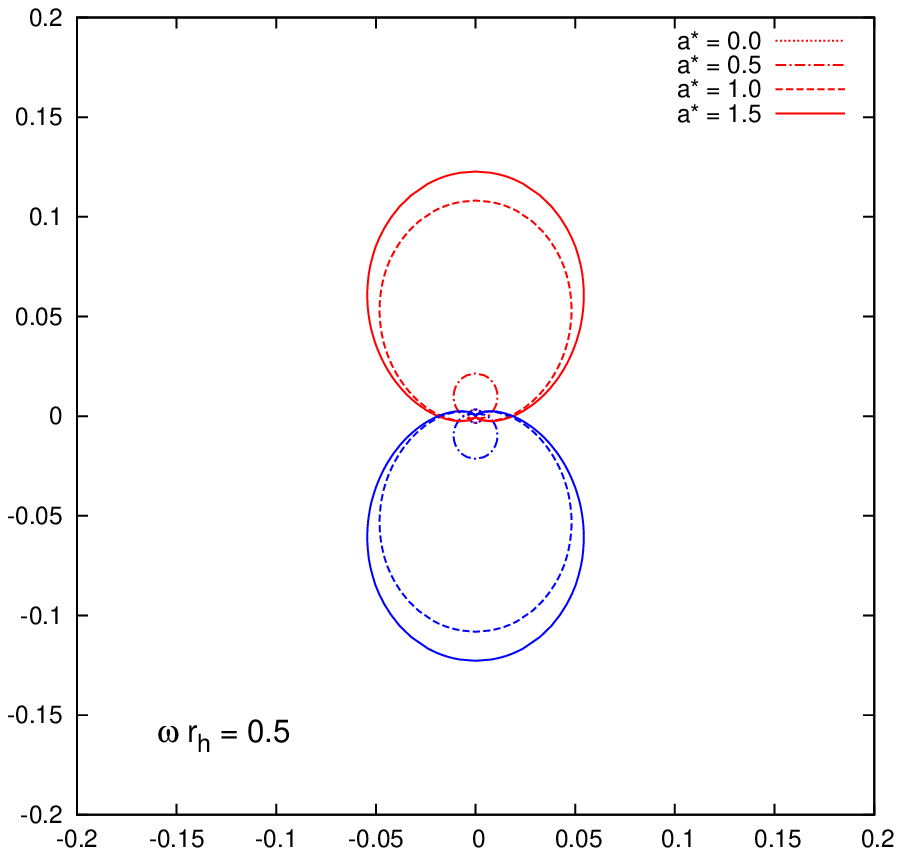}} \hspace*{-0.5cm}
{\includegraphics[scale=0.58]{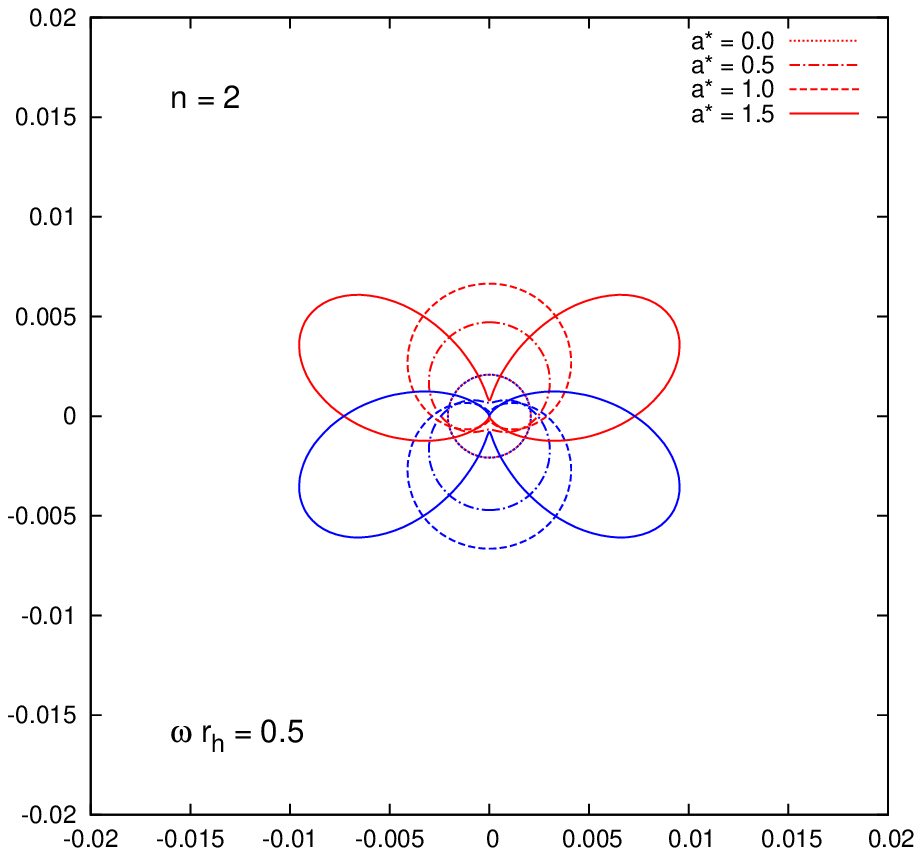}}
\caption{\label{ang-s-polar} Polar plots with the angular distribution of
gauge bosons (left plot) and fermions (right plot) for $\omega r_h=0.5$,
$n=6$ and variable $a_*$ \cite{CDKW3}. Red (upper) curves correspond to positive
helicity particles and blue (lower) curves to negative helicity particles.}
\label{degeneracy}
\end{center}
\end{figure}

As we mentioned above, one of the two factors that determine the angular distribution
of the emitted particles is the spin-rotation coupling that is dominant at the
low-energy regime. If we then focus on this part of the energy spectrum,
and consider the emission of fermions and gauge bosons, we find the
behaviour depicted in Figure \ref{degeneracy} \cite{CDKW3}: the gauge bosons
(left plot) are aligned parallel or anti-parallel to the rotation axis of the
black hole \cite{CDKW3}, while the fermions (right plot) have an angle
of emission that depends on the value of the angular momentum of
the black hole \cite{CDKW3,FST}. The aforementioned pattern is in fact
independent of the value $n$ of extra dimensions. Therefore, by observing
the angles of emission of gauge bosons and fermions in a low-energy channel,
one could in principle determine the orientation of the rotation axis and
the angular momentum, respectively, of the black hole. The diferential
fluxes (\ref{eq:particleflux}--\ref{eq:angmomflux}) could then be used to
determine the value of $n$, too.

The results depicted in Figure \ref{degeneracy} were found by numerically
integrating both the radial and angular equations for gauge bosons and
fermions. However, since the behaviour found above takes place in a
low-energy channel, one could attempt to use analytic methods to solve
both equations and derive the angular emission pattern. As we have already
discussed, the radial equation (\ref{eq:radialbrane}) has been solved analytically
for all species of brane particles \cite{IOP1,CEKT2,CEKT3}. The
angular equation (\ref{eq:angularbrane}) also admits an analytic solution in the
form of an infinite power-series \cite{Leaver}. In \cite{KP2}, a constraint
was thus derived that semi-analytically determines the angle of maximum emission
for different types of fields; the angular emission pattern of scalars, gauge
bosons and fermions was then found, for a wide range of values of the
angular-momentum parameter $a_*$ and energy $\omega$ of the emitted particle.


\subsection{Emission of massless fields in the bulk}

Having studied the emission of Hawking radiation on the brane, i.e. the part
of the emitted energy that a brane observer could potentially detect, we now
turn to the emission of Hawking radiation in the bulk, i.e. the part of the
emitted energy that would literally go missing. Brane observers have no
access to the bulk, nevertheless, we need to study the different types of
emission and estimate the amount of energy that is channeled
in the bulk. Since Standard Model particles are constrained to live on the
brane, the only degrees of freedom allowed to propagate in the bulk are
scalars and gravitons. In what follows, we review the existing results in
the literature for the corresponding radiation spectra.

We start with the emission of scalar fields: the set of equations,
for the more complex rotating phase, are given in
(\ref{eq:radialscalarbulk}--\ref{eq:angularscalarbulk}), while
the ones for the spherically-symmetric phase follow by setting $a=0$.
These equations have been solved, for both phases, by employing either
analytical \cite{KMR1, FS0, CEKT4} or numerical \cite{Casals:2008pq,HK, JP05} methods.
For the analytical approach, based on the same approximation method as in
the case of brane emission, we need the angular eigenvalue $\lambda_\Lambda$
in an analytical form: for the spherically-symmetric phase, this is known and given by \cite{Mullerbook}
\begin{equation}
\lambda_\Lambda=\ell (\ell+n+1),
\end{equation}
while for the rotating phase
there is again a power-series expression \cite{Eigenvalues1, Eigenvalues2}.
The analytical results thus derived for the grey-body factor may be used to
describe the effect of Hawking radiation in the bulk very accurately
in the low-energy regime and, at times, even in the intermediate-energy regime.

However, the complete spectra may be derived only through numerical integration.
For the spherically-symmetric case, the angular equation (\ref{eq:angularscalarbulk})
contains again no new information for the radiation process - it is only
the radial equation (\ref{eq:radialscalarbulk}), with $a=0$, that needs to be
numerically integrated. For the rotating phase, the angular equation
(\ref{eq:angularscalarbulk}) is integrated first to provide the
eigenvalue $\lambda_\Lambda$ - note that, in the absence of an observer in the
bulk, there is no motivation for the study of the angular distribution of the
emitted particles. The radial equation (\ref{eq:radialscalarbulk}) is integrated
next to determine the grey-body factors and, subsequently, the emitted fluxes.

\begin{figure}
\begin{center}
\mbox{ \hspace*{-4.0cm}
\includegraphics[scale=0.55]{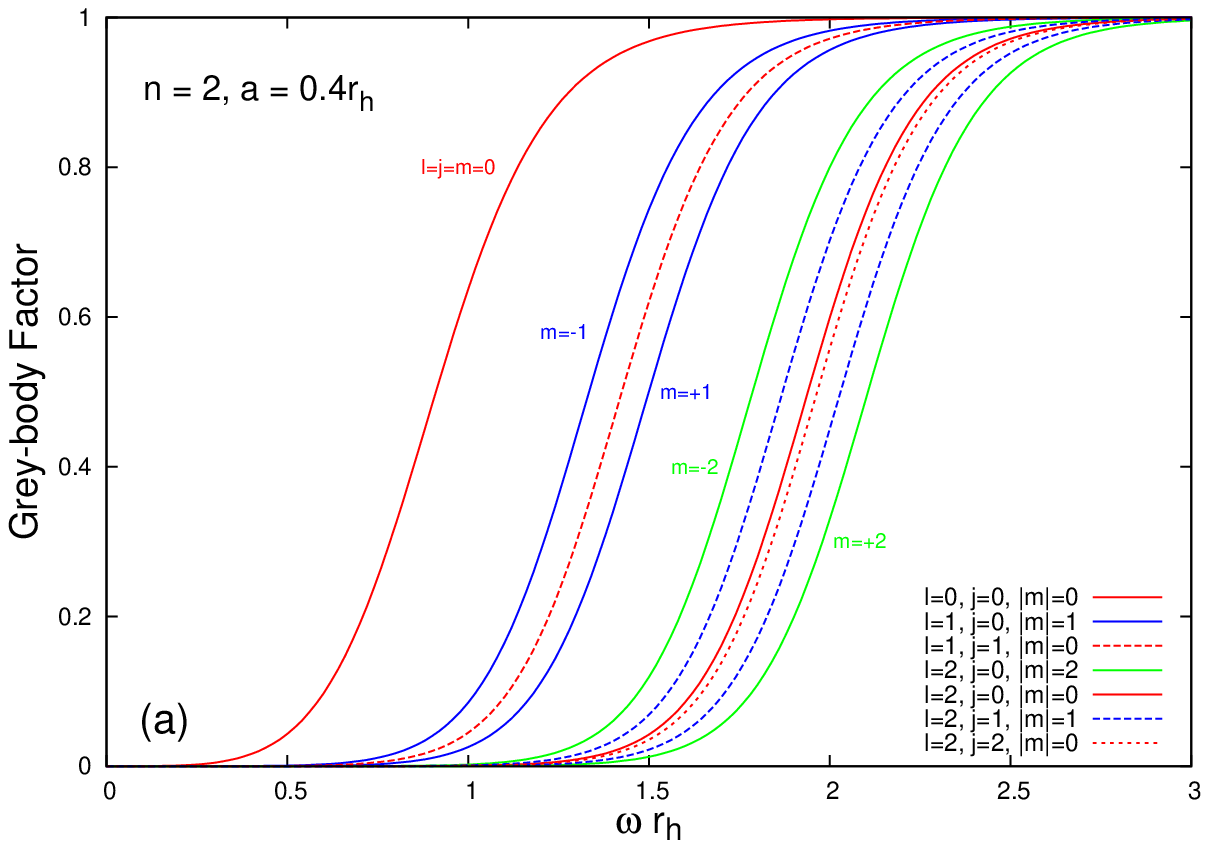}} \hspace*{-0.7cm}
{\includegraphics[scale=0.53]{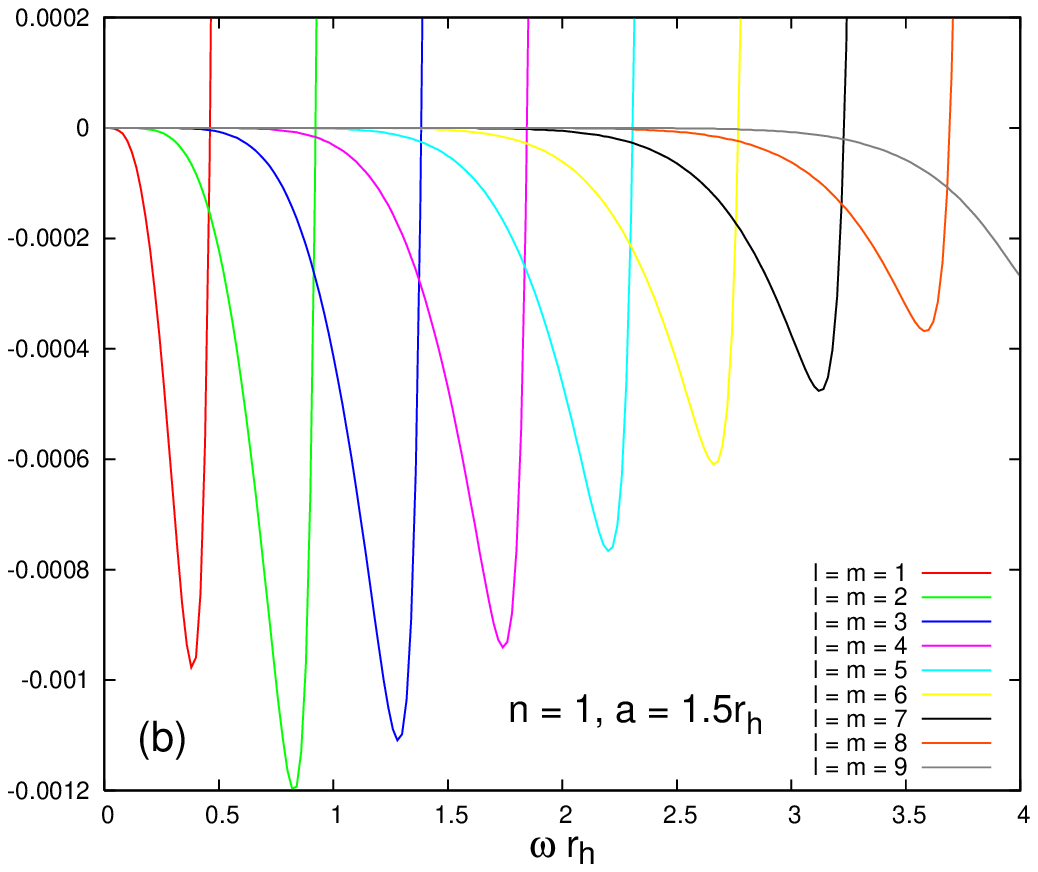}}
\caption{Grey-body factors for a bulk scalar field: {\bf (a)} complete results for
various modes $(j,\ell,m)$, for $n=2$ and $a_*=0.4$, and {\bf (b)}
the super-radiant regime for various co-rotating modes $\ell=m$, for $n=1$ and
$a_*=1.5$ \cite{Casals:2008pq}.}
   \label{grey-s0-bulk}
\end{center}
\end{figure}

We will focus on the presentation of results for the rotating phase, since the
dependence on the number of extra dimensions $n$ will
become manifest when we fix the value of the angular-momentum parameter $a_*$.
In Figure \ref{grey-s0-bulk}(a) \cite{Casals:2008pq}, we present exact numerical
results for the grey-body factor for a scalar field emitted in the bulk by a
six-dimensional black hole with $a_*=0.4$: the different curves correspond to
various modes characterised by the set of $(j,\ell,m)$ quantum numbers and
show a hierarchical splitting first on $\ell$, then on $m$ and finally on $j$.
Figure \ref{grey-s0-bulk}(b)
\cite{Casals:2008pq} depicts the behaviour of the grey-body factor for a bulk scalar
field, in the background of a black hole with $n=1$ and $a_*=1.5$, in the
super-radiant regime, $\omega < m\,\Omega_H$: as in the case of brane emission,
the super-radiance effect is most important for the maximally co-rotating modes
$\ell=m$ and $j=0$, and for low values of $n$.

\begin{figure}[b!]
  \begin{center}
  \mbox{ \hspace*{-14.0cm}
\includegraphics[scale=0.55]{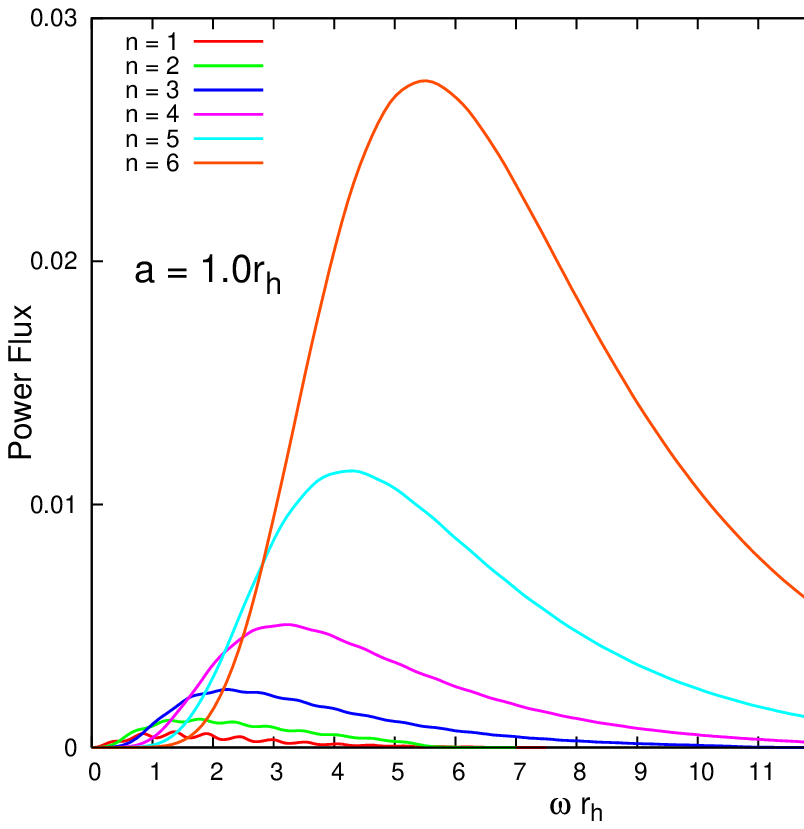}} \hspace*{2cm}
{\includegraphics[scale=0.55]{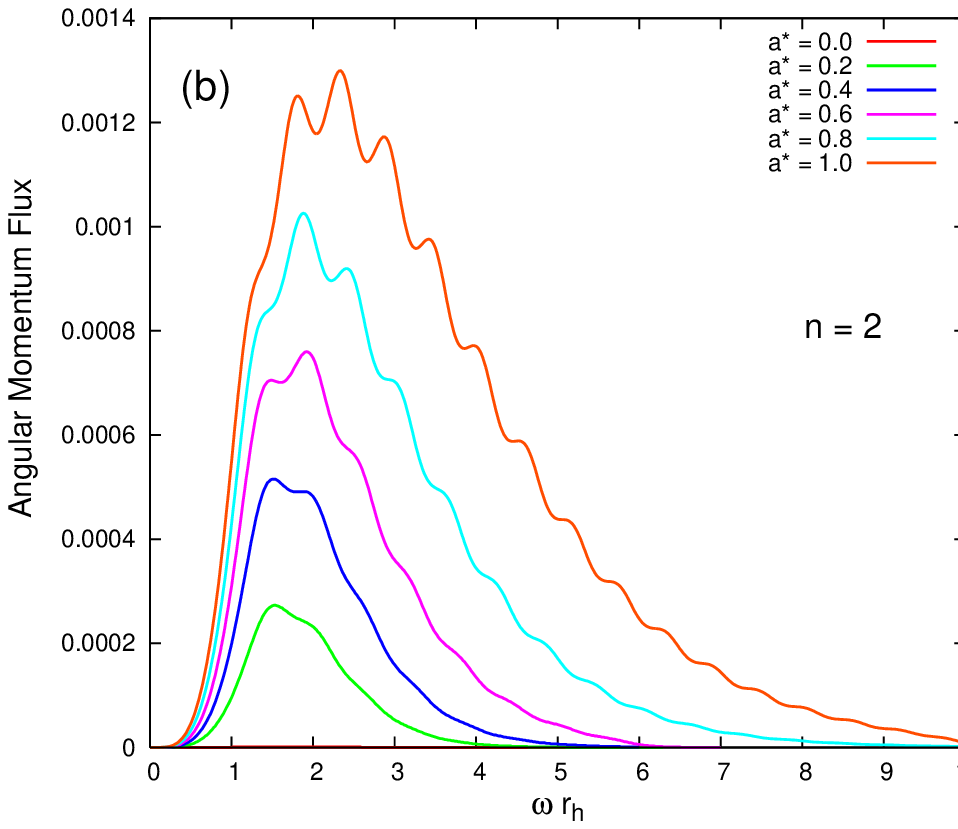}}
\caption{ {\bf (a)} Energy emission rate for bulk scalar fields from a black hole
with $a_*=1$ and variable $n$, and {\bf (b)} angular-momentum emission rate for bulk
scalar fields from a black hole with $n=2$ and variable $a_*$ \cite{Casals:2008pq}.}
   \label{Power-ang-bulk.eps}
  \end{center}
\end{figure}

\begin{figure}[t!]
  \begin{center}
  \mbox{ \hspace*{-12.5cm}
\includegraphics[scale=0.6]{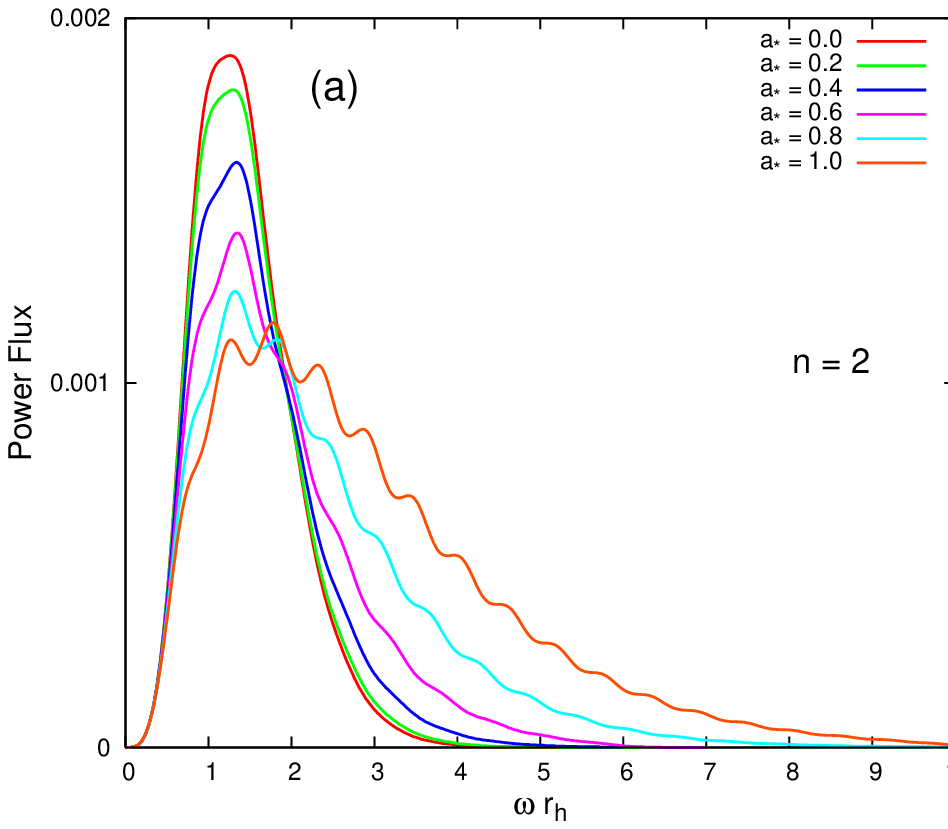}} \hspace*{-0.2cm}
{\includegraphics[scale=0.6]{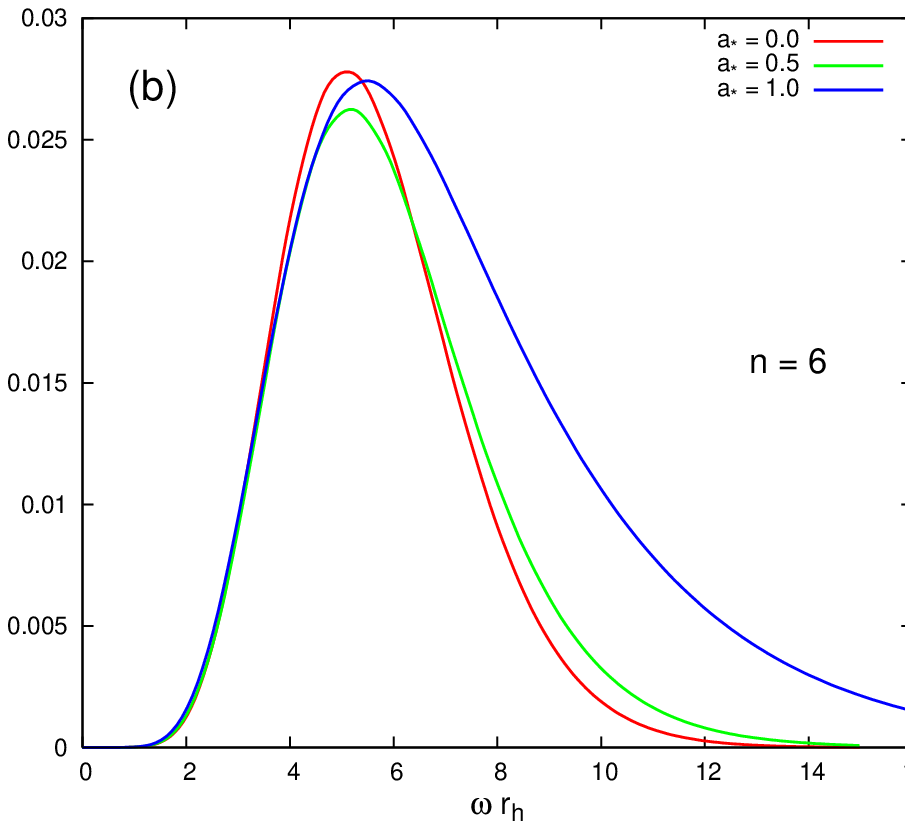}}
\caption{ Energy emission rates for scalar fields in the bulk from a
black hole with {\bf (a)} $n=2$, and {\bf (b)} $n=6$, and variable $a_*$ \cite{Casals:2008pq}.}
   \label{Power-bulk_a.eps}
  \end{center}
\end{figure}

In Figure \ref{Power-ang-bulk.eps}(a), we depict the differential energy
emission rate for bulk scalar fields \cite{Casals:2008pq}, for fixed angular-momentum
parameter $a_*=1$, and variable $n$. The power flux shows a significant
overall enhancement as $n$ increases, with a small suppression at the low-energy
regime and a shift of the peak of the curve towards larger energies - this
behaviour is also identical to the one observed in the case of a
spherically-symmetric black hole emitting scalar particles in the bulk
\cite{HK}. The particle and angular-momentum fluxes were also found to have the
same behaviour. The angular-momentum emission rates, presented in
Figure \ref{Power-ang-bulk.eps}(b) for $n=2$, show a significant enhancement over
the whole energy regime, as $a_*$ increases.
The power and particle fluxes, on the other hand, have a more particular profile:
for low values of $n$, the emission curves are also shifted to the high-energy
regime, as $a_*$ increases, but their peak values are significantly suppressed,
as shown in Figure \ref{Power-bulk_a.eps}(a); for large values of $n$, the energy
and particle emission curves remain almost unchanged as $a_*$ increases, apart
from a small enhancement in the amount of emission at the high-energy regime;
see Figure \ref{Power-bulk_a.eps}(b).

We now turn to the emission of gravitons in the bulk by a higher-dimensional
black hole. For a spherically-symmetric black hole, the analysis is now complete.
The radial equation (\ref{eq:gravbulk}), obeyed by all three types of
gravitational perturbations - scalar, vector and tensor ones - has been solved
analytically both in the low \cite{Creek:2006ia} and intermediate-energy
regime \cite{CNS}. The low-energy analysis \cite{Creek:2006ia} revealed
that the graviton spectra exhibit the same behaviour as bulk
scalar fields \cite{HK}, i.e. a suppression at the low-energy regime and a
shift of the emission curve towards higher energies, as $n$ increased. The
same analysis showed that, at the lower part of the energy spectrum, the
emission of gravitons is negligible compared to that of bulk scalar fields,
however that was expected to change at higher energies where higher partial
modes would dominate.

That expectation was indeed proved right by the exact numerical analysis
performed in \cite{CCG, CCG1, Park}. It was thus demonstrated that the graviton
radiation spectra were strongly enhanced at the higher part of the spectrum
and for large values of $n$. The latter was largely due to the fact that
the degeneracy factors of the graviton states (\ref{eq:gravdegen1})
increase rapidly with both $n$ and $\ell$ with the higher modes dominating at the
upper part of the spectrum. For example, for a moderate value of $\ell$,
i.e. $\ell=5$, the number of graviton states, as $n$ varies from 1 to 6,
increases by a factor of $10^4$ \cite{Creek:2006ia}. An interesting twist is that
the tensor graviton modes, the most negligible degrees of freedom at the
low-energy regime \cite{Creek:2006ia}, proliferate as $n$ increases.
Overall, it is found \cite{CCG,CCG1} that, as $n$ reaches the value 7, a
spherically-symmetric black hole emits 35 times more energy in the bulk
in the form of gravitons than in any other particle on the brane.

\begin{figure*}[t!]
  \begin{center} \hspace*{-0.5cm}
\mbox{\includegraphics[width =.53\textwidth]{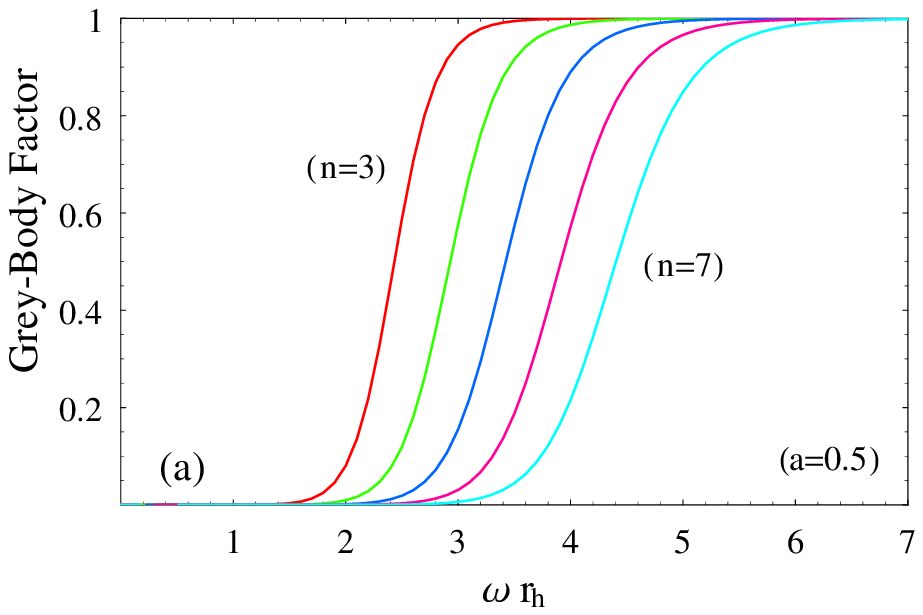}}\hspace*{-0.7cm}
  {\includegraphics[width =.53\textwidth]{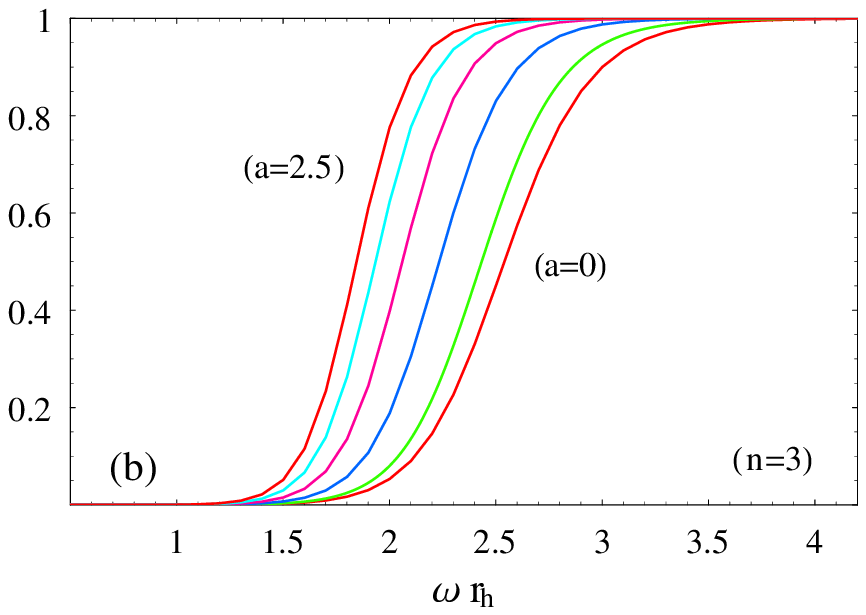}}
    \caption{\label{fig2-gravs} Grey-body factors for tensor-type graviton
	modes $(\ell=j=2,m=0$) for: {\bf (a)} $a_*=0.5$ and variable $n$, and
	{\bf (b)} $n=3$ and variable $a_*$ \cite{Kanti:2009sn}.}
  \end{center}
\end{figure*}

\begin{figure*}[b!]
  \begin{center}
 \includegraphics[width = 0.5\textwidth]{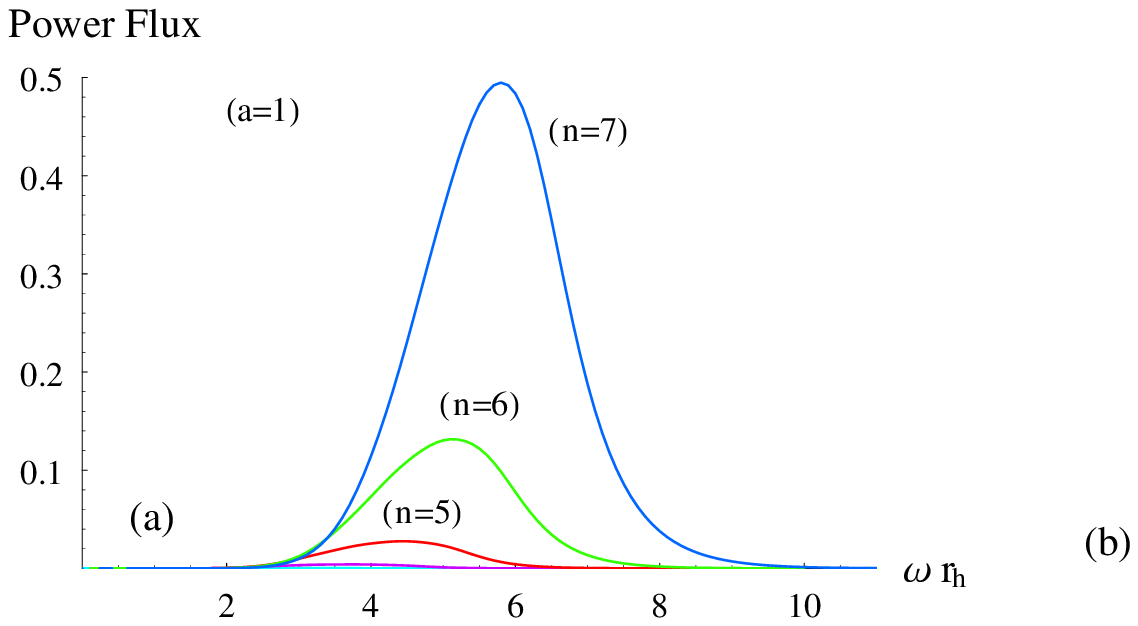}\hspace*{-0.2cm}
 \includegraphics[width = 0.5\textwidth]{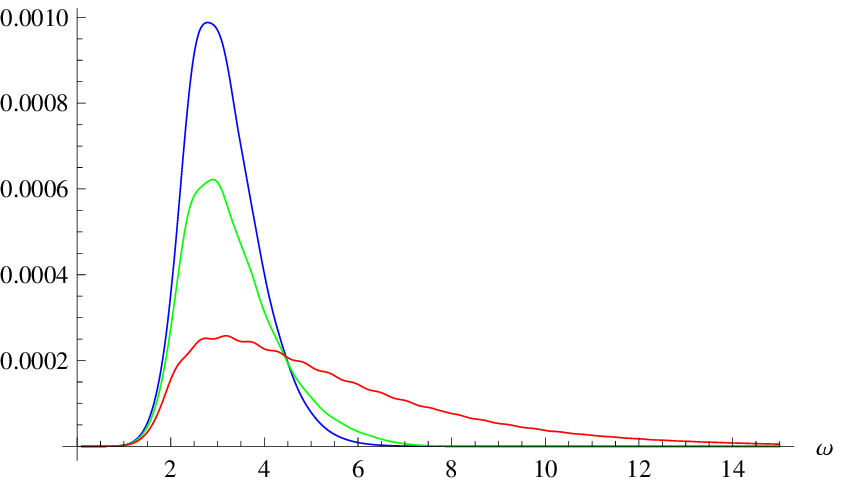}
    \caption{\label{fig6-gravs} Energy emission rate for tensor-type gravitons
	{\bf (a)} for $a_*=1$ and variable $n$, and {\bf (b)} for $n=3$ and
	$a_*=0, 0.5,1.2$ (from top to bottom) \cite{Kanti:2009sn}.}
  \end{center}
\end{figure*}

For the case of the simply-rotating black hole (\ref{eq:MyersPerry}), only the
equations for tensor-type perturbations have been derived, and these are identical
to the ones for a bulk scalar field
(\ref{eq:radialscalarbulk}--\ref{eq:angularscalarbulk}) apart from the allowed
values of the angular-momentum number $l$ (i.e. $\ell \geq 2$ instead of $\ell \geq 0$).
These equations were solved both analytically and numerically in \cite{Kanti:2009sn},
with the two sets shown to agree remarkably well even up to the intermediate-energy
regime. In Fig. \ref{fig2-gravs}(a,b), we present exact numerical results for the
grey-body factor of the lowest tensor-type mode ($\ell=j=2, m=0$), in terms
of $n$ and $a_*$, respectively \cite{Kanti:2009sn}. We observe that, as the
number of extra dimensions $n$ increases, the grey-body factor is suppressed
over the whole energy regime -- a similar suppression is exhibited also by
bulk scalar fields \cite{Casals:2008pq,HK,CEKT4}. On the other hand, the grey-body
factor is strongly enhanced with the angular momentum of the black hole.
Turning to the fluxes, the indicative case of the energy emission rate is
presented in Figures \ref{fig6-gravs}(a,b) \cite{Kanti:2009sn}, in terms
again of $n$ and $a_*$. Similarly to the case of bulk scalars, the graviton
emission rates are significantly enhanced with $n$, while the emission is
suppressed at the low and intermediate-energy regimes and enhanced at the
high-energy regime as $a_*$ increases (for large $n$, the suppression at the low and
intermediate-energy regimes is replaced again by a mild dependence on $a_*$).

%
%

\subsection{Energy balance between the brane and the bulk}

The question of the energy balance between the brane and bulk emission channels
is an important one, not only for its theoretical interest but also for practical
purposes: if we determine what fraction of the available energy of the black hole
is lost in the bulk, we will know how much remains for emission on the brane,
and thus how likely the observation of the Hawking radiation effect will be
for the brane observer.

The study of the higher-dimensional spherically-symmetric black hole is, as we saw,
now complete with the exact spectra for all types of brane and bulk particles
being determined. In \cite{Casals:2008pq,HK}, exact numerical analyses were performed
in order to compare the scalar emissivity of the black hole in the brane and
bulk channels. In the first row of Table \ref{emiss-scalars} we display some
indicative values of the proportion of the total power emitted in the bulk
by a non-rotating black hole \cite{Casals:2008pq,HK}: we see that the bulk scalar
channel is always subdominant to the brane one, but not necessarily negligible.
Although bulk and brane scalar fields ``see'' the same black-hole temperature,
their grey-body factors behave differently with $n$, leading to the emission of
more energetic, but significantly fewer bulk scalar fields compared to the
brane ones.

For the brane-localised fermions and gauge bosons and the bulk gravitons, it is
again the relative behaviour of their grey-body factors, but also of their degeneracy
factors, that will determine the result.
Regarding the latter, we have already discussed the rapid proliferation, with
$n$, of the gravitons in the bulk, however, the large number of Standard Model degrees
of freedom living on the brane must also be taken into account. When all the
above are implemented in the analysis, it is found \cite{CCG,CCG1} that the brane
channel is the most dominant during the spherically-symmetric phase of the
black hole, a result that agrees with an early analytical argument \cite{EHM}.

\begin{table}[b!]
\caption{Total proportion of scalar power emitted in the bulk by a black-hole  \cite{Casals:2008pq}}
\begin{center}
{\begin{tabular}{ccccccccc} \hline
$n$ & 1 & 2 & 3 & 4 & 5 & 6 \\ \hline
\hspace*{0.15cm} $a_*=0.0$ \hspace*{0.2cm}& \hspace*{0.2cm} 28.3$\%$
\hspace*{0.15cm} & \hspace*{0.2cm} 19.9$\%$ \hspace*{0.2cm}& \hspace*{0.2cm} 17.9$\%$
\hspace*{0.15cm} & \hspace*{0.2cm} 19.6$\%$ \hspace*{0.2cm} & \hspace*{0.2cm}24.8$\%$
\hspace*{0.15cm} & \hspace*{0.2cm} 34.0$\%$ \hspace*{0.2cm}\\
\hspace*{0.15cm} $a_*=0.5$ \hspace*{0.2cm}& 20.9$\%$ & 13.5$\%$ & 11.8$\%$ & 13.0$\%$ & 16.7$\%$
& 24.0$\%$ \\
\hspace*{0.15cm} $a_*=1.0$ \hspace*{0.2cm}& 12.5$\%$ & 7.1$\%$ & 6.2$\%$ & 6.8$\%$ & 9.1$\%$
& 14.7$\%$ \\\hline
\end{tabular}} \end{center} \label{emiss-scalars}
\end{table}

The same question of the bulk-to-brane energy balance needs to be posed also
for the rotating phase. The relative scalar emissivity can again be derived
since the radiation spectra for bulk and brane emission are known
\cite{Casals:2008pq,Duffy:2005ns,HK2}. The entries of Table \ref{emiss-scalars}
reveal that, not only is the bulk scalar channel the subdominant one
also during the rotating phase, but the proportion of the energy emitted
in the bulk reduces with the angular-momentum of the black hole.
This is due to the fact that the enhancement of the grey-body factor with
$a_*$ is not as large for bulk scalar emission as it is for the brane one.

Since we still lack the complete emission spectra for all types of gravitational
perturbations in the bulk, the question of the energy balance between the brane
and bulk channels for the rotating phase remains open. In \cite{Kanti:2009sn},
the total emissivity of tensor-type gravitons was compared to the one for scalar
bulk emission. It was found that the energy emitted in the bulk, for small values of
$n$, in the form of tensor modes is less than 1\% of the scalar emissivity  but it
becomes of the order of 25\%, for $n=5$ and for the indicative value of $a_*=1$.
Recalling that the tensor modes were the dominant gravitational ones in the bulk
in the case of the spherically-symmetric phase, we may conclude
that, for low values of $n$, the brane channel wins the energy-balance
contest in the rotating phase, too. Whether the same situation holds for
large values of $n$ (or larger values of $a_*$) will be decided only when the
exact emission spectra for vector and scalar-type graviton modes are  found.

\subsection{Additional effects in Hawking radiation}
\label{sec:additional}

\begin{figure}
\begin{center}
\hspace*{-0.5cm}
\mbox{\includegraphics[scale=0.7]{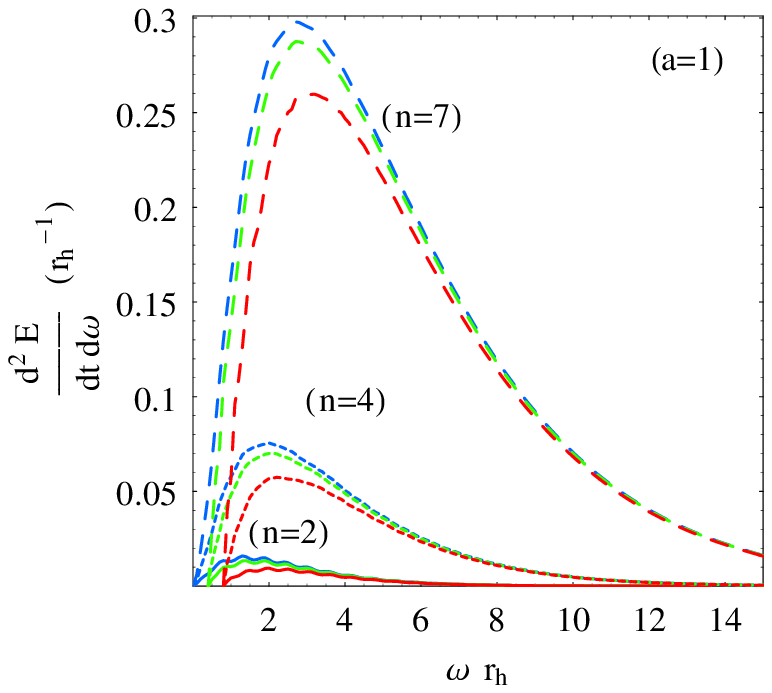}} \hspace*{0.1cm}
{\includegraphics[scale=0.35]{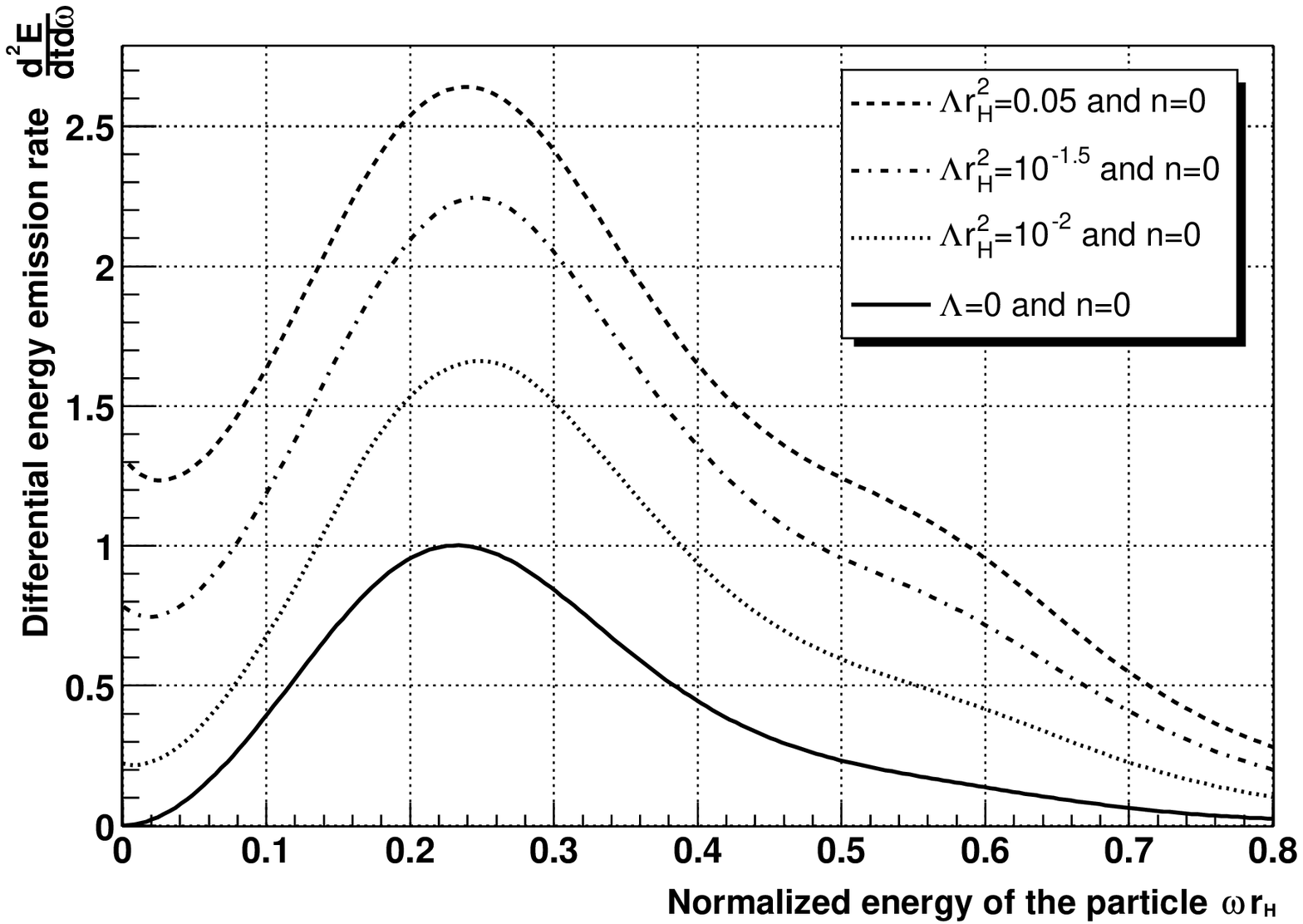}}
\caption{\label{kan-massive} Energy emission rates for scalar fields on the
brane: {\bf (a)} with $a_*=1$, and a mass $m_\Phi=0,0.4,0.8$
(from top to bottom in each set of curves) \cite{KP1}, and {\bf (b)} from a
Schwarzschild-de Sitter black hole with various values of $\Lambda$ \cite{KGB}.}
\label{Mass-Lambda}
\end{center}
\end{figure}

We now discuss some further aspects of the Hawking
radiation emission process. In all the studies mentioned so far, the
different types of particles emitted by the black hole were assumed
for simplicity to be massless. The presence of the mass, however, is expected
to cause a suppression of the grey-body factors since the emission of a
massive field demands more energy, and thus it is less likely to happen.
The effect of the mass on the radiation spectra was studied in
\cite{Sampaio:2009ra, KP1, Sampaio}, for scalar fields  emitted by a
higher-dimensional rotating black hole, and in \cite{HSW, WSH}, for vector
fields, both transverse and longitudinal modes, on a $D$-dimensional
Schwarzschild background. It was demonstrated that the suppression is
indeed more prominent the larger the mass of the emitted field, as depicted
in Figure \ref{Mass-Lambda}(a) \cite{KP1}. Although the brane channel
remains the dominant one, in \cite{KP1}, it was shown that the presence
of the mass enhances the bulk-over-brane energy ratio up to a factor of 34\%.

The effect of the charge of the emitted particles was studied in \cite{Sampaio,
WSH, Jung}. In \cite{Jung}, a spherically-symmetric Reissner-Nordstr\"om black
hole was considered, and it was shown that the charge of a scalar particle,
similarly to its mass, causes a suppression in the bulk and brane emission
spectra and enhances the bulk-to-brane emissivity. In \cite{WSH}, the case of
a charged vector field was studied, and an inverted charge splitting effect
as well as the analogue of a superradiance effect were observed in the radiation
spectra. The case of a higher-dimensional rotating black hole with a
brane-localised Maxwell field was considered in \cite{Sampaio}. Although
only brane emission of scalars and fermions was studied, an interesting
effect was found: the radiation process exhibits a charging-up phase at
the low-energy regime (where the black hole emits particles with an opposite
charge compared to its own) while a phase of discharging dominates at
higher energies.

In addition to particle properties, parameters that characterize the black-hole
space-time background may also affect the radiation spectra and subsequently the
bulk-to-brane energy balance. The cosmological constant $\Lambda$, that may be
present in the part of the space-time where the black hole is formed, is one of
these parameters. In \cite{KGB}, the radiation spectra for scalar fields emitted
by a higher-dimensional Schwarzschild-de Sitter black hole were studied.
It was found that both bulk and brane emission rates are enhanced as the
value of $\Lambda$ increases, with the spectrum exhibiting a novel feature, i.e.  a
non-vanishing emission rate at the lowest part of the energy regime (see,
Figure \ref{Mass-Lambda}(b)). The presence of the cosmological constant
was shown to increase the bulk-to-brane relative emissivity. In the case of
a Kerr-de Sitter black hole \cite{DCCN}, a similar enhancement is found
when the Bousso-Hawking definition of the black hole temperature is used,
whereas the use of the Hawking definition leads instead to a small decrease.

Higher-order curvature terms, such as the Gauss-Bonnet term, may be also taken
into account during the decay of the black hole. Their effect on the emission
spectra strongly depends on the mass of the black hole, the value of the coupling
constant, and the type and energy of the emitted particle: for scalar fields,
they can tilt the bulk-to-brane energy ratio in favour of the bulk \cite{BGK},
while, for gravitons, they can cause the suppression of their emission by many
orders of magnitude leading to an increased life-time of the black hole
\cite{KZ}. Another counter-example of the claim that black holes
radiate mainly on the brane appeared in \cite{CCDN} where it was shown
that, in the context of a supersymmetric split-fermion model, the bulk fermion
emission dominates over the brane one for $n > 1$.
On the other hand, including non-commutative geometry-inspired corrections to the black hole metric
significantly lowers the black hole temperature, and the bulk emission is greatly suppressed \cite{Nicolini:2011nz}.


\section{Hawking radiation from black holes in the RS model}
\label{sec:BBHrad}

Given the absence of an analytical solution describing a five-dimensional
asymptotically AdS black hole localised on the brane, no exact results
may be presented for the Hawking radiation process from such a black hole.
Nevertheless, a number of partial or approximate results have appeared
in the literature that may shed light on some aspects of this process.

In the RS-II model, the AdS length scale $\ell_{AdS}$ plays the role of
the effective size of the fifth dimension. On the other hand, the size
of the black hole is given by the horizon radius $r_h$. If
$r_h \ll \ell_{AdS}$, the black hole is insensitive to the warping along
the fifth dimension, and it resembles a black hole in a five-dimensional flat
space-time whose line-element is described by either the Myers-Perry (\ref{eq:MyersPerry}) or the
Tangherlini solution. In this limit, it shares all the properties of
black holes in a higher-dimensional flat space-time. For example, if we
assume for simplicity that the black hole is non-rotating, its
horizon radius-mass relation and temperature will be given by
\begin{equation}
r_h=\sqrt{\frac{8}{3\pi}}\,\frac{1}{M_5}\,\left(\frac{M}{M_5}\right)^{1/2},
\qquad T_{BH}=\frac{1}{2\pi r_h}\,,
\end{equation}
where $M$ and $M_5$ are the black-hole mass and the fundamental Planck scale,
respectively. The above expressions follow readily from the corresponding
$(4+n)$-dimensional ones \cite{Argyres} after setting $n=1$. Compared to a
four-dimensional black hole of the same mass, for which
$r_h=2M/M_P^2$ and $T_{BH}=1/4\pi r_h$, a small brane-world black hole
has a larger horizon radius and, thus, a smaller temperature. Therefore,
its evaporation rate will be significantly suppressed leading to a longer
lifetime \cite{Guedens}
\begin{equation}
\frac{t_{\rm evap}(5D)}{t_{\rm evap}(4D)} \sim
\left(\frac{\ell_{AdS}}{r_h(5D)}\right)^2\,,
\end{equation}
the smaller the horizon radius of the five-dimensional black hole is compared
to $\ell_{AdS}$. The above, combined with an increased accretion rate for
a brane-world black hole being formed in the early universe, radically change the
mass spectrum of the primordial black holes that are expected to decay today
\cite{Guedens, Majumdar, Sendouda}.

As the horizon radius increases, the black hole starts to perceive the warping
of the fifth dimension and its properties are correspondingly modified.
Unfortunately, only numerical solutions are available in the literature in
this regime. In \cite{Kudoh-thermo}, the thermodynamic properties of such
a five-dimensional black-hole solution localised on the brane were studied.
It was demonstrated that, for a small value of the horizon radius, the
properties of the black hole match those of a five-dimensional Schwarzschild-AdS
black hole. For a large value of the horizon radius, they tend to the ones of
a four-dimensional Schwarzschild black hole, while in the intermediate regime
they remain quite distinct from these two limits.

For $r_h \gg \ell_{AdS}$, we expect the brane-world black hole to obtain a peculiar
shape: while there is no limit on its size along the brane directions, in the bulk it
cannot extend at distances larger than $\ell_{AdS}$ due to the localisation of gravity.
Thus, a large brane-world black hole has a ``pancake'' shape. Although it is genuinely a
higher-dimensional object, it extends so little along the extra dimension and so much
along the usual three that we expect it to be effectively four-dimensional and thus to
resemble the Schwarzschild solution. That is, if it is allowed by the theory to exist.
Even then, it is not exactly Schwarzschild.

As it was briefly mentioned in section 3.1, in reality, the projected line-element on the
brane is not a vacuum solution. A four-dimensional observer would independently write the
effective Einstein's equations on the brane as $G_{\mu\nu}=8\pi G_N\,T_{\mu\nu}$.
According to the AdS/CFT correspondence \cite{HHR}, this $T_{\mu\nu}$ is the expectation
value of the stress-energy tensor operator for the CFT modes for a suitably chosen
vacuum state. The correspondence dictates that the number of CFT modes living on
the brane is given by $N \sim (\ell_{AdS}/\ell_P)^2$ (where $\ell _{P}$ is the Planck length) and thus, for
$\ell_{AdS} \gg  \ell_P$, this number is expected to be very large.
In \cite{Tanaka}, it was then argued that the presence of such a large number of
fields in the vicinity of the black hole will greatly enhance its evaporation
rate resulting in a large-mass black hole having a life-time given by \cite{EGK}
\begin{equation}
t_{\rm evap}=116 \times \left(\frac{1\,{\rm mm}}{\ell_{AdS}}\right)^2
\left(\frac{M}{M_\odot}\right)^3\,{\rm years}\,.
\end{equation}
For such a rapidly evaporating black hole, no static solution can be found to
describe it even approximately. Since the line-element on the brane is merely
the projection of the five-dimensional one, the bulk solution could not be static
either, but rather must describe an evaporating black hole on the brane while
remaining classical. In \cite{Tanaka}, a ``classical evaporation'' process
was proposed: instabilities of the solution along the bulk cause the deformation
of the horizon radius and its elongation towards the AdS boundary; for its
total horizon area to remain constant, the area of the intersection of the
black hole with the brane will have to decrease, and that will be viewed by
a four-dimensional observer as an evaporation process.

The authors of \cite{Fitzpatrick} counter-argued that the above scenario
assumes the involvement of all CFT brane modes in the evaporation process.
But if the CFT on the brane is strongly-coupled, as the AdS/CFT correspondence
dictates, would all modes be available to interact with the four-dimensional
black hole? Most likely not, therefore one
should not expect important quantum corrections to the four-dimensional
line-element. In support of this, in \cite{Zegers} a Schwarzschild-AdS$_4$
black string, proven to be stable, was considered and the question was asked:
why is this five-dimensional classical solution quantum-corrected on the
brane (the same question for the black string of \cite{CHR} loses part
of its validity due to its instability)? The stress-energy tensor for the
boundary field theory was computed and shown to lead to a mere renormalization
of the effective cosmological constant, and not to a radiation term.
In addition, in \cite{Kaloper} it was recently suggested that a low-energy
theory, such as the CFT theory on the brane, needs to be carefully
UV-completed before predictions for the IR behaviour of the theory are
made. When this is properly done, it follows that the emission of CFT modes
in the RS-II model is significantly suppressed - in fact, it was argued that,
at distances $L \leq N\,\ell_P$, low-energy CFT does not even exist.

The numerical nature of the solutions found in \cite{FW, Page, Page1},  describing
both small and large black holes in the RS-II model, does not allow us to
analytically study the effect that the brane tension or the bulk cosmological
constant have on the evaporation process. In the context of a six-dimensional
model, a solution was found \cite{KK} that described a black hole localised
on a codimension-two brane  with tension. The solution was asymptotically flat,
and not AdS as in the RS-II model, and it had a peculiar horizon radius-mass
relation: the closer the brane tension to the fundamental scale $M_6$,
the larger the horizon radius is. It was shown \cite{Dai} that an increase
in the brane tension simultaneously decreases the black-hole temperature
and increases the potential barrier, thus causing the suppression of both
bulk and brane emission rates.


\section{Conclusions}
\label{sec:conc}

In this chapter we have reviewed some key aspects of our understanding of Hawking radiation from higher-dimensional black holes.
One motivation for this study, apart from its intrinsic theoretical interest, is the exciting possibility of observing Hawking radiation from
microscopic higher-dimensional black holes produced in high-energy collisions at either the LHC or in cosmic rays.

We began with a discussion of the quantum-field-theoretic foundations of Hawking radiation, from its original derivation for a quantum field on
a dynamical space-time representing a black hole formed by gravitational collapse, to its modelling using the Unruh state on an eternal black hole
geometry.
After a brief review of classical black hole geometries in both the ADD and RS brane-worlds, most of this chapter has been devoted to the Hawking
emission from these black holes.

For black holes in the ADD brane-world, for which we have an exact space-time metric, we gave an overview of the formalism used to describe the emission of
massless particles both on the brane and in the bulk.
We then reviewed a selection of results on the nature of the emission, including the important question of how much Hawking radiation escapes into the bulk
and is therefore inaccessible to a brane-localised observer.
Except for the emission of scalar- and vector-like graviton modes from a rotating black hole, our understanding of the emission of massless particles is essentially complete.
We also briefly summarized some features of the Hawking radiation when other effects, such as the mass and charge of the emitted particles or a cosmological constant, are included.

In the RS brane-world, the lack of an exact classical metric for a black hole localized on the brane implies a scarcity of precise results on the nature of the Hawking emission from such black holes.  We have therefore just briefly outlined some of the work in the literature in this case.

Our review has revealed that Hawking radiation from higher-dimensional black holes is more complicated than from four-dimensional black holes, with many features depending on the number of extra dimensions and the angular momentum of the black hole.
Many of the results we have outlined in this chapter have been incorporated into simulations of black hole events at the LHC \cite{Dai:2007ki,Frost:2009cf}.
Hawking radiation from black holes at the LHC characteristically involves large numbers of energetic particles, the details of the distribution and nature of the particles depending on the number of extra dimensions, the mass of the black hole and its angular momentum.
Searches for these types of events have been made at the LHC, with no evidence to date for black holes \cite{Aad:2011bw,Aad:2012ic,Aad:2013lna,Khachatryan:2010wx,Chatrchyan:2012taa,Chatrchyan:2013xva}.
This non-observation of black hole events sets lower bounds on the energy scale of quantum gravity.
With the LHC planned to run at higher energies and searches for high-energy black hole events in cosmic rays,  accurate modelling of Hawking radiation
from higher-dimensional black holes continues to be essential for experimental probes of quantum gravity.

\begin{acknowledgement}
The work of E.W. is supported by the Lancaster-Manchester-Sheffield
Consortium for Fundamental Physics under STFC Grant ST/J000418/1.
The work of P.K. has been co-financed by the European Union (European
Social Fund - ESF) and Greek national funds through the Operational Program
``Education and Lifelong Learning'' of the National Strategic Reference
Framework (NSRF) - Research Funding Program: ``ARISTEIA. Investing in the
society of knowledge through the European Social Fund''. Part of this work
was supported by the COST Actions MP0905 ``Black Holes in a Violent Universe''
and MP1210 ``The String Theory Universe''.
\end{acknowledgement}

\end{document}